\documentclass[a4paper,fleqn]{cas-dc}
\usepackage{siunitx}
\usepackage[authoryear]{natbib}

\usepackage{graphicx}
\usepackage[labelformat=empty, position=top]{subcaption}
\usepackage[export]{adjustbox}
\usepackage{xcolor}

\definecolor{mediumseagreen}{rgb}{0.231, 0.620, 0.392}
\definecolor{deepskyblue}{rgb}{0.118, 0.565, 1}
\definecolor{purpleline}{rgb}{0.627, 0.125, 0.941}

\def\tsc#1{\csdef{#1}{\textsc{\lowercase{#1}}\xspace}}
\tsc{WGM}
\tsc{QE}
\tsc{EP}
\tsc{PMS}
\tsc{BEC}
\tsc{DE}

\begin{document}
\let\WriteBookmarks\relax
\def\floatpagepagefraction{1}
\def\textpagefraction{.001}

\shorttitle{}

\shortauthors{H. Chen et~al.}

\title [mode = title]{Spatio-Temporal Modeling of Surface Water Quality Distribution in California (1956-2023)}                      


%
\author[1]{Houlin Chen}[orcid=0009-0002-2611-9387,]

\fnmark[1]



\credit{Conceptualization of this study, Methodology, Software}

\author[2, 3]{Meredith Franklin}[orcid=0000-0003-3802-8829]

\cormark[1]

\affiliation[1]{organization={Faculty of Arts and Science},
    addressline={University of Toronto}, 
    city={Toronto},
    state={Ontario},
    country={Canada}}

\affiliation[2]{organization={Department of Statistical Sciences and School of the Environment},
    addressline={University of Toronto}, 
    city={Toronto},
    state={Ontario},
    country={Canada}}
    
\affiliation[3]{organization={Department of Population and Public Health Sciences},
    addressline={University of Southern California}, 
    city={Los Angeles},
    state={CA},
    country={USA}}

\cortext[cor1]{Corresponding author at: Department of Statistical Sciences, University of Toronto, 700 University Ave, Toronto, Ontario, Canada.}
  
\fntext[fn1]{Faculty of Arts and Science, University of Toronto, 100 St George St, Toronto, Ontario, Canada.}

\nonumnote{\textit{E-mail addresses}:
houlin.chen@mail.utoronto.ca (H. Chen), meredith.franklin@utoronto.ca (M. Franklin).}

\begin{abstract}
Surface water quality has a direct impact on public health, ecosystems, and agriculture, in addition to being an important indicator of the overall health of the environment. California's diverse climate, extensive coastline, and varied topography lead to distinct spatial and temporal patterns in surface water. This study offers a comprehensive assessment of these patterns by leveraging around 70 years of data, taking into account climate zones and geographical types. We analyzed surface water quality indicators, including pH, dissolved oxygen, specific conductance, and water temperature, based on field results from approximately 5,000 water quality stations in California Water Quality Data (CWQD). Machine learning (ML) models were developed to establish relationships between spatial and temporal variables, climate zones, geographical types, and water quality indicators. Applying these models to spatially interpolate and temporally predict the four water quality indicators over California for the next 50 years, the research results indicate an uneven distribution of water quality indicators in California, suggesting the presence of potential pollution zones, seawater erosion, and effects of climate change.
\end{abstract}




\maketitle

\section{Introduction}

The quality of surface water is an integral factor in various aspects of human and ecosystem life. It is affected by a diverse range of elements, from microbial content (\cite{khadra2022systematic}) to industrial effluents (\cite{kaur2010physico}), to Earth's water cycle (\cite{sauve2021circular}), all playing a significant role in determining the quality of surface water. A notable illustration of a significant water quality issue is the nuclear wastewater leak in Fukushima, Japan, in 2011 (\cite{buesseler2011impacts}). The effects of the leak reached the United States coastline in just three years due to atmospheric pressure and ocean currents, with repercussions on water quality that are projected to extend for over 30 years (\cite{rossi2013multi}). Water plays a crucial role in human well-being and health, making the investigation and prediction of water quality trends an essential area of research (\cite{li2019drinking}).

\begin{figure}
    \centering
    \includegraphics[width=1\linewidth]{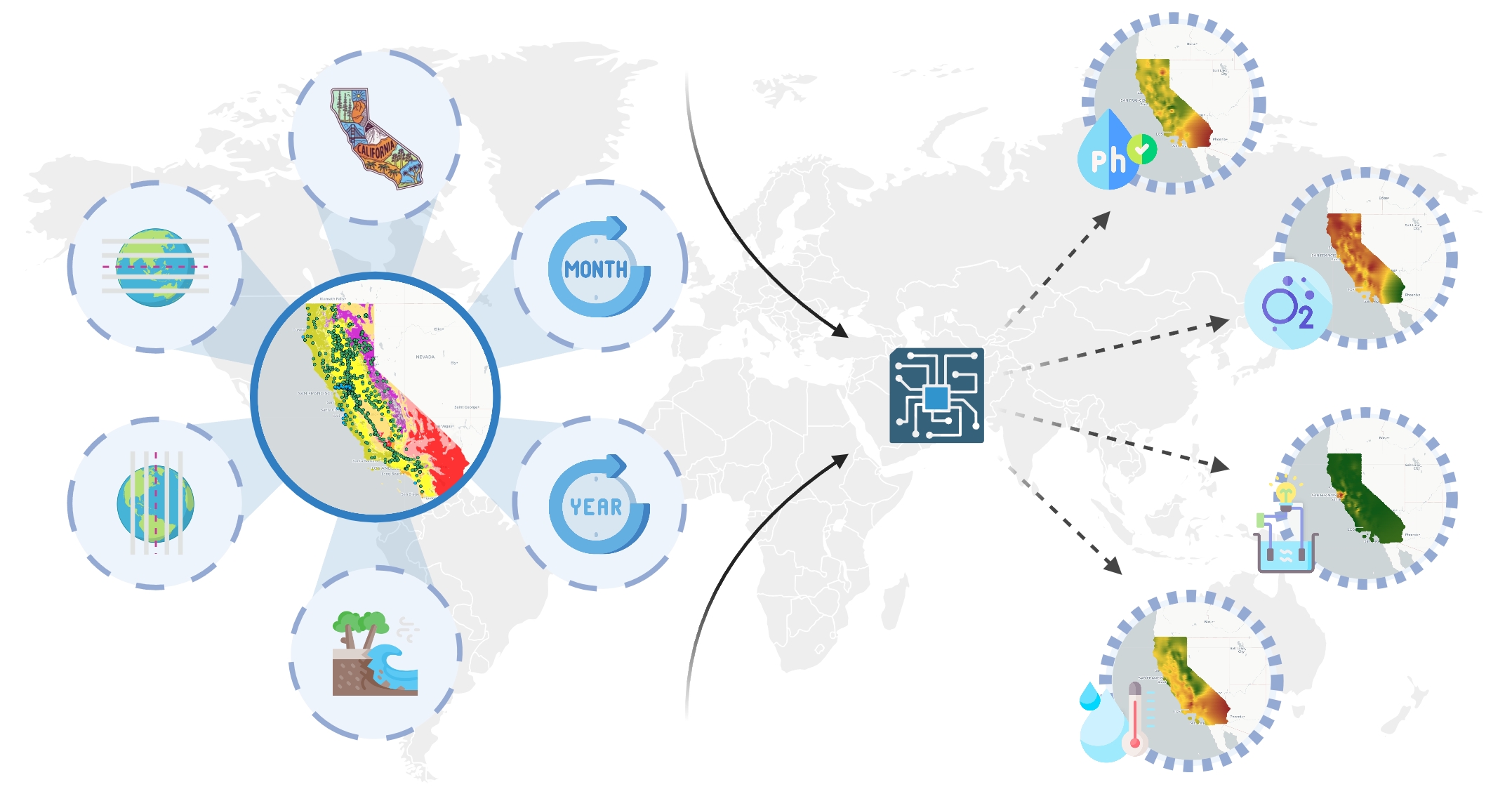}
    \caption{System diagram of the spatio-temporal modeling in this paper.}
    \label{fig:abstract}
\end{figure}

The World Health Organization (WHO) outlines in its Guidelines for Drinking-Water Quality (GDWQ) (\cite{world2022guidelines}) that assessing water quality encompasses the evaluation of physical, chemical, and biological parameters (\cite{icaga2007fuzzy, zhu2022review}). This includes concentrations of harmful substances such as nitrates, the presence of organisms like E. coli, and levels of radioactive materials. For example, the GDWQ explicitly links the concentration of contaminants and microbiological content to various water safety levels, drawing attention to the relationship between the proportion of E. coli negative samples and the quality of drinking water systems for different population sizes (refer to Table 5.2 in GDWQ). It is crucial to note that these water quality indicators are defined under specific environmental conditions. For example, the WHO identifies a pH range of 6.5-8.5 at 25°C as ideal for eco-friendly and health-promoting drinking water. However, an increase in water temperature could lead to a decrease in pH levels. Consequently, the distribution of these water quality indicators is susceptible to climatic conditions, temperature variations, and geographical features, posing significant challenges to researchers attempting to estimate and predict these interrelated indicators.


Water quality indices (WQIs) are generated by aggregating multiple water-related measures available from temporally and spatially-varying water quality datasets into a single value \cite{Uddin2021, chidiac2023comprehensive}. As an example in the United States, a commonly used WQI standard is the National Sanitation Foundation (NSF) index, which consists of nine variables and is widely accepted and applied (\cite{brown1970water, noori2019critical}). As for the examples of WQI application, \cite{jha2020assessing} proposed a WQI model based on fuzzy geographical information systems to perform large-scale analyses of groundwater quality, and \cite{kim2019quantitative} established a water quality risk index (WQRI) for Korea based on a relationship between water quality variables and drought indices determined by a kernel density estimator. WQIs are useful as they provide a straightforward way for the public and policy makers to interpret complex data, however, they often do not account for regional or local features, nor do they explicitly address spatial and temporal variability, which may differ for each of the component variables that make up a WQI. This implies that when our focus is solely on a specific water quality parameter, WQIs may not provide an intuitive representation. Furthermore, as WQIs are artificially formulated and adjusted based on regional and temporal variations, they may not objectively and accurately reflect the long-term changes and future trends of water quality.

California, a highly populated state characterized by its diverse climates and micro-climates, as well as varied geographical features, has been the subject of significant environmental research. It hosts three major climate types - arid, temperate, and cold (\cite{kauffman2003climate}), and also experiences seawater erosion on its coastal freshwater system. Water quality concerns in California have predominantly been about nitrate contamination from agricultural activities in the Central Valley such as fertilizer application \cite{pennino2020}. Elevated concentrations of nitrate have been associated with adverse birth outcomes \cite{sherris2021} and thyroid cancer \cite{tariqi2021water}. Other markers of water quality, including those examined in this study, are important to monitor to ensure California's water is safe for human consumption and as a habitat for aquatic life, which indirectly affects health through the food chain. In a United States Geological Survey (USGS) report, pH measured at 1,337 wells over a 20-year period from 1993 to 2014 in California's Central Valley was modelled and mapped to provide an understanding of water quality conditions at domestic and public supply drinking water zones \cite{rosecrans2017predicted}. Another USGS report analyzed specific conductance and water temperature at eight stations in the San Francisco Bay Area from 1990 to 2015, found that both parameters reached record highs in the region in 2015 (\cite{work2017record}). However, the existing literature on water quality in California lacks comprehensive spatio-temporal modeling that accounts for the long-term assessment and integration of multiple indicators, and often overlooks the diverse range of climate zones and geographical types (\cite{ficklin2013watershed, li2019beach}).



In this paper, we conduct a spatio-temporal analysis of water quality indicators (pH, dissolved oxygen, specific conductance, and water temperature) collected in California over the past 70 years, spanning from 1956 to 2023. We establish relationships between spatio-temporal variables and these indicators using regression and machine learning (ML) models (Fig.~\ref{fig:abstract}). Through this approach we reveal potential interactions among water quality indicators, climate zones, and geographical types. The trained spatio-temporal models were applied to spatially interpolate the water quality indicators over the state, revealing potential areas of pollution or poor ecological health. Furthermore, we used the models to project trends in the indicators for the next 50 years. This allowed us to examine how ecological and climate changes in California have impacted surface water, suggesting future environmental protection policies. California can be considered a microcosm for water quality analysis under most geographical types and major climate conditions, offering guidance for water quality analysis in regions with similar climatic conditions or topography.

\begin{figure*}
    \centering
    \includegraphics[width=1\linewidth]{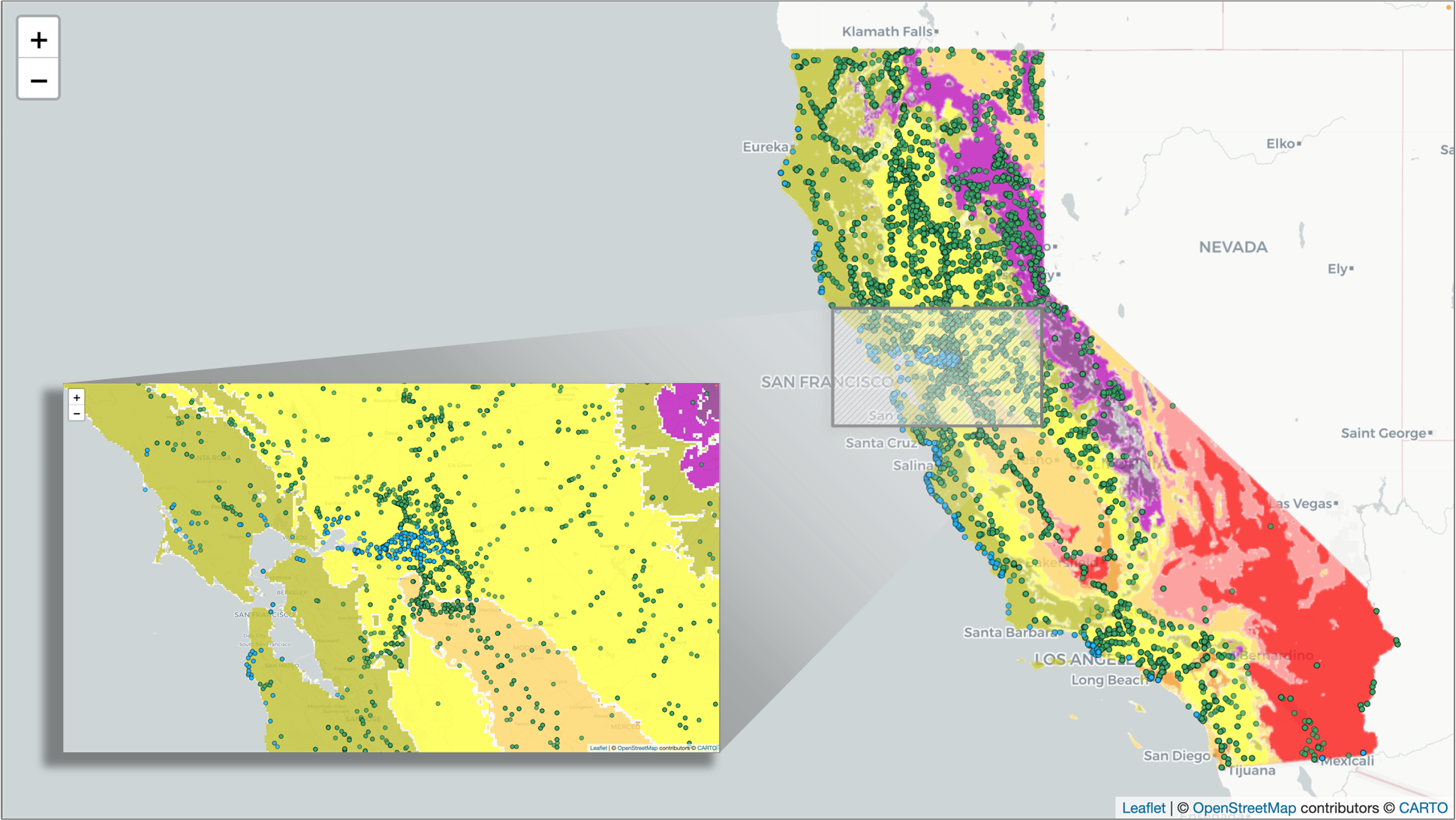}
    \caption{Distribution of water quality stations in California. Points on the map represent stations, with \textcolor{mediumseagreen}{Inland} and \textcolor{deepskyblue}{Coastal} in Geographical Type, respectively. The varied background colors delineate different climate zones based on the Köppen climate classification.}
    \label{fig.station distribution}
\end{figure*}

\begin{figure*}
    \centering
    \includegraphics[width=1\linewidth]{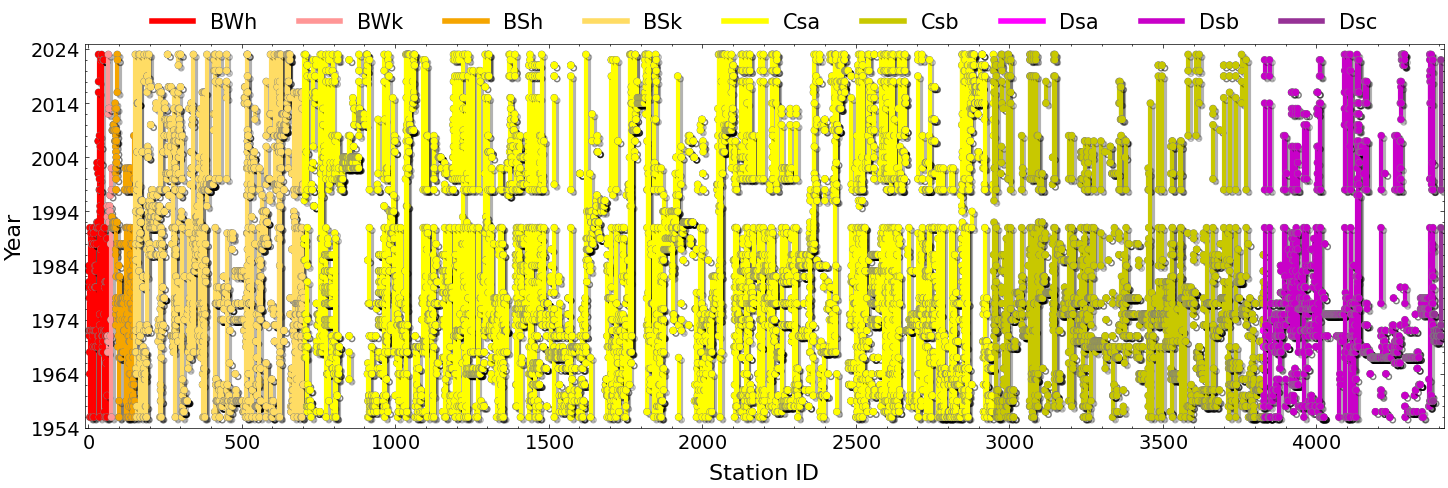}
    \caption{Lifelines of water quality stations in California. Each line represents the duration of existence for a station, with its color indicating the climate zone of the station's location (Fig.~\ref{fig.station distribution}). See sub-climates abbrevitions in Section~\ref{koppen}.}
    \label{fig.lifeline}
\end{figure*}

\begin{table*}[h]
\scriptsize
\caption{A preview of the preprocessed dataset CWQD used in our analysis, representing water quality in various stations of California (\cite{CaliforniaWaterQualityData}).}\label{tb: example data}
\begin{tabular}{@{}lcccccccccccc@{}}
\toprule
\textbf{Station} & \textbf{Latitude} & \textbf{Longitude} & \textbf{County} & \textbf{Year} & \textbf{Month} & \textbf{pH} & \textbf{Dissolved} & \textbf{Specific} & \textbf{Water} & \textbf{Climate} & \textbf{Geographical}\\
\textbf{ID} & (\(^\circ\)) & (\(^\circ\)) &  &  &  & (pH Units) & \textbf{Oxygen} & \textbf{Conductance} & \textbf{Temperature} & \textbf{Zone} & \textbf{Type}\\
 &  &  &  &  &  &  & (mg/L) & (\(\mu S/\text{cm}@25 ^\circ \text{C}\)) & (\(^\circ \text{C}\)) &  & \\
\midrule
1 & 37.8019 & -121.6203 & Alameda & 1975 & 1 & 6.9 & 11.1 & 415.0 & 8.9 & BSk & Inland\\
2 & 37.5636 & -121.6866 & Alameda & 1983
 & 9 & 7.3 & 5.3 & 407.0 & 14.5 & Csb & Inland\\
3 & 37.6147 & -121.7458 & Alameda & 2021 & 10 & 7.0 & 9.9 & 567.0 & 17.3 & Csb & Inland\\
4 & 37.8149 & -121.5527 & Alameda & 1994 & 7 & 7.9 & 7.0 & 551.0 & 26.2 & BSk & Inland\\
5 & 37.6183 & -121.7494 & Alameda & 2023 & 5 & 8.3 & 6.5 & 332.0 & 12.6 & Csa & Inland\\
\bottomrule
\end{tabular}
\end{table*}

\section{Materials and Methods}
\subsection{California Water Quality Dataset}
The California Water Quality Dataset (CWQD) was sourced from the California Department of Water Resources (\cite{CaliforniaWaterQualityData}). It encapsulates a comprehensive amalgamation of both field and laboratory results, underpinned by various physical and chemical parameters. From a temporal perspective, the CWQD documents a staggering 110 years of water quality indices, starting in May 1913, and it continues to evolve with real-time updates (Fig.~\ref{fig.lifeline}). Throughout its lifecourse, the CWQD has included a total of 29,229 water quality monitoring stations across the state (Fig.~\ref{fig.station distribution}). This is a vast spatial representation, as California spans a longitudinal breadth of \(10.3^\circ\) and a latitudinal extent of \(9.47^\circ\). This expansive scope covers diverse terrains, extending from deserts to islands and highlands to basins. 

The CWQD provides three distinct classifications of water based on varying depths: surface water, under-surface water, and groundwater. The field results in the CWQD include parameters such as pH, Dissolved Oxygen, Specific Conductance, and Water Temperature (Table~\ref{tb: example data}). In terms of lab results, it encapsulate indices that mandate more extended testing periods and intricate procedural steps, such as Dissolved Calcium, Dissolved Chloride, and Total Hardness, among others. In this paper, we conduct a thorough analysis of 5,080 water quality monitoring stations for surface water parameters, spanning from January 1956 to July 2023, within the scope of real-time field results from the CWQD. Surface water quality data have been continuously collected at these 5,000 over this time span, with most of them having samples collected four times per month.  

The variables available in CWQD are all important indicators of surface water quality. pH, serving as a crucial measure of water's acidity or alkalinity, directly reflects the solubility of toxic metals, the degree of eutrophication, and the disruption of physiological functions in aquatic organisms. Dissolved Oxygen is a pivotal standard for biological survival, signifies one of the most foundational criteria for assessing whether a considerable biological congregation exists in rivers, lakes, etc. Specific Conductance, through a more cost-effective and expeditious method, indirectly mirrors ion concentrations of water quality indices. On the contrary, it is challenging to measure Dissolved Calcium in the field. Water Temperature provides an indirect reflection of water quality, wherein alterations in temperature can signify shifts in the chemical, biological, and physical characteristics of a body of water, consequently influencing the health and stability of the entire ecosystem.

\subsection{Köppen Climate and Geographical Types}
\label{koppen}
Climate zones are inextricably linked to water quality due to the pronounced impact different temperatures and precipitation levels exert on water quality indices (\cite{barbieri2021climate}). For example, varying temperatures can influence the solubility of substances in water, while differing precipitation levels may affect the concentration of dissolved substances by altering water flow and levels. Consequently, we employ the Köppen climate classification (\cite{beck2018present}) to attribute nuanced climate labels to each station in California, as depicted in Fig.~\ref{fig.station distribution}. The Köppen climate classification is a globally acknowledged and utilized climate categorization system, extensively applied in various disciplines such as meteorology, climatology, and environmental science, to facilitate tasks like regional climate categorization, biodiversity studies, and climatic impact analyses. The Köppen climate classification encapsulates five primary climate types, namely tropical (A), arid (B), temperate (C), cold (D), and polar (E). Each of these major classifications is further refined into a total of 30 sub-climates based on nuanced differences in precipitation, temperature variations, and geographical features. The sub-climates are differentiated by factors such as the type of seasonal precipitation and the degree of heat, exemplified by classifications such as BSk, which stands for arid (B), steppe (S), and cold (k), and Csb, which represents temperate (C), dry summer (s), and warm summer (b). When formulating features or variables for our models, we employ the broader classifications, incorporating only the pertinent three, as the climatic specifics of California do not involve tropical (A) or polar (E) classifications. Applying the finely segmented sub-climates to California, the state is delineated into nine distinct sub-climates. Specifically, the nine sub-climates include arid, desert, hot (BWh); arid, desert, cold (BWk); arid, steppe, hot (BSh); arid, steppe, cold (BSk); temperate, dry summer, hot summer (Csa); temperate, dry summer, warm summer (Csb); cold, dry summer, hot summer (Dsa); cold, dry summer, warm summer (Dsb); and cold, dry summer, cold summer (Dsc) (Fig.~\ref{fig.lifeline}).

The composition and concentration of constituents vary significantly between seawater and inland waters, mainly due to the high salinity and pollutants found in seawater. The intrusion of seawater into coastal areas and its potential impact on inland surface waters can significantly disrupt our assessments of inland water quality. In turn, this could affect the proposed method for evaluating inland water quality through surface water assessments, which is crucial for supporting smart agriculture and other applications. Extensive literature review reveals that there is no conclusive agreement on the distance, intensity, and depth of seawater intrusion into inland surface waters (\cite{abd2022saltwater, mentaschi2018global, hussain2019management}). Consequently, we propose categorizing stations located within eight kilometers of the coastline as Coastal stations, and those situated more than eight kilometers away as Inland stations (Fig.~\ref{fig.station distribution}). 

\subsection{Experimental Setup}

In the data pre-processing phase, we first rectified errors within the entire dataset by correcting or removing them, addressing issues such as non-negative longitude and stations with identical latitude and longitude coordinates but different station IDs, and eliminating all observations with NA. Subsequently, removed outliers, discarding data points that fall outside the 95\% confidence interval. After conducting these preprocessing steps, we had a total of 64,185 samples, which we then partitioned into training and testing subsets at an 8:2 ratio, using the training set to develop the model and the unseen test set to evaluate its accuracy. For the evaluation of model performance, we employed Root Mean Square Error (RMSE) and the coefficient of determination ($R^2$) as our principal metrics. RMSE provides insights into the discrepancies between the actual values in the dataset and the predictions from the model, while $R^2$ quantifies the explanatory power of the model.

\subsection{Analysis}
Machine learning, with its ability to perform nonlinear parameter fitting, has seen widespread adoption in the estimation of spatio-temporal variables to capture complex patterns and interactions within data (\cite{tahmasebi2020machine, yuan2023three, yuan2023passive}). We consider six regression models as potential candidates for data fitting. The Linear Model (LM) is a method of predicting a response variable, such as a expected water quality indicator, through a weighted sum of predictor variables (e.g., month, year, longitude, latitude, etc.) and a constant term (a.k.a. the intercept) plus an error term. Random Forest (RF) is an ensemble technique that produces a more reliable result by combining the predictions of multiple Decision Trees (DTs). Each tree is trained on a random subset of the data and predicts the expected water quality indicator based on predictor variables. The Gaussian Process (GP) estimates mean and variance parameters by fitting the response and predictor variables with a covariance function (\cite{guinness2021gaussian}). Prediction of new data is based on observations and its covariance structure. Support Vector Machine (SVM) finds a hyperplane in a high-dimensional space that is as far away from observations of different categories as possible to achieve regression. The Generalized Additive Model (GAM) introduces nonlinear functions on the basis of LM, allowing complex relationships between response variables and predictor variables transformed through nonlinear functions, providing greater flexibility and accuracy for prediction (\cite{hastie2017generalized}). Extreme Gradient Boosting (XGBoost) is also an ensemble method that combines the predictions of multiple DTs to generate a final prediction (\cite{chen2016xgboost}). In addition to RF, XGBoost adds new trees at each step based on the error of the previous tree and uses gradient boosting techniques to guide this process to gradually optimize prediction performance. Model performance is optimized through extensive parameter tuning.

\begin{table*}
    \caption{The RMSE of each water quality indicator predicted by various ML methods on the test set. S-T and V-D refer to Spatio-Temporal Estimation (Section \ref{s-t}) and Variable-Dependent Inference (Section \ref{v-d}), respectively. The lowest RMSE for each water quality indicator is highlighted in \textbf{bold}.}
    \label{tab:rmse}
    \centering
    \begin{tabular}{@{}lllllllll@{}} \toprule  
         & \multicolumn{2}{l}{\textbf{pH}}&  \multicolumn{2}{l}{\textbf{Dissolved Oxygen}}&  \multicolumn{2}{l}{\textbf{Specific Conductance}}&  \multicolumn{2}{l}{\textbf{Water Temperature}}\\ 
        & \multicolumn{2}{l}{\textbf{(pH Units)}}&  \multicolumn{2}{l}{\textbf{(mg/L)}}&  \multicolumn{2}{l}{\textbf{(\(\mu S/\text{cm}@25 ^\circ \text{C}\)
        )}}&  \multicolumn{2}{l}{\textbf{(\(^\circ \text{C}\)
        )}}\\ \cmidrule{1-9}
         \textbf{Method} &  \textbf{S-T} &  \textbf{V-D} &  \textbf{S-T} &  \textbf{V-D} &  \textbf{S-T} &  \textbf{V-D} &  \textbf{S-T} &  \textbf{V-D}\\ \midrule
         \textbf{LM}& 0.548& 0.498& 1.989& 1.913& 1285.151& 405.445& 5.086&4.516\\
         \textbf{RF}& 0.408& 0.378& 1.362& 1.452& 631.505& 257.467& 1.918&1.859\\
         \textbf{GP}&  0.446&  0.465&  1.483&  1.856&  733.691&  346.392&   2.234&2.757\\
         \textbf{SVM}&  0.495&  0.428&  1.718&  1.649&  1273.048&  380.100&  2.524&2.221\\
         \textbf{GAM}&  0.478&  0.432&  1.686&  1.715&  961.122&  348.542&  2.445&2.306\\
         \textbf{XGBoost}&  \textbf{0.402}&  \textbf{0.376}&  \textbf{1.355}&  \textbf{1.380}&  \textbf{601.385}&  \textbf{247.900}&  \textbf{1.838}&\textbf{1.738}\\ \bottomrule 
    \end{tabular}
\end{table*}

\begin{table*}
    \caption{The $R^2$ of each water quality indicator predicted by various ML methods on the test set. S-T and V-D refer to Spatio-Temporal Estimation (Section \ref{s-t}) and Variable-Dependent Inference (Section \ref{v-d}), respectively. The highest $R^2$ for each water quality indicator is highlighted in \textbf{bold}. }
    \label{tab:r2}
    \centering
    \begin{tabular}{@{}lllllllll@{}} \toprule  
         & \multicolumn{2}{l}{\textbf{pH}}&  \multicolumn{2}{l}{\textbf{Dissolved Oxygen}}&  \multicolumn{2}{l}{\textbf{Specific Conductance}}&  \multicolumn{2}{l}{\textbf{Water Temperature}}\\  \cmidrule{1-9}
         \textbf{Method} &  \textbf{S-T} &  \textbf{V-D} &  \textbf{S-T} &  \textbf{V-D} &  \textbf{S-T} &  \textbf{V-D} &  \textbf{S-T} &  \textbf{V-D}\\ \midrule
         \textbf{LM}&  0.085&  0.207&  0.059&  0.230&  0.051&  0.210&  0.161&0.339\\
         \textbf{RF}&  0.493&  0.542&  0.559&  0.557&  0.771&  0.682&  0.881&0.900\\
         \textbf{GP}&  0.393&  0.309&  0.476&  0.275&  0.691&  0.425&  0.838&0.757\\
         \textbf{SVM}& 0.254& 0.415& 0.298& 0.428& 0.068& 0.306& 0.793&0.842\\
         \textbf{GAM}&  0.302&  0.402&  0.324&  0.381&  0.469&  0.417&  0.806&0.830\\
         \textbf{XGBoost}&  \textbf{0.507}&  \textbf{0.548}&  \textbf{0.563}&  \textbf{0.599}&  \textbf{0.792}&  \textbf{0.705}&  \textbf{0.890}&\textbf{0.903}\\ \bottomrule
    \end{tabular}
\end{table*}

\section{Results}

\subsection{Spatio-Temporal Estimation}
\label{s-t}
We explore the relationship between various spatio-temporal variables (i.e., Month, Year, Latitude, and Longitude) and individual water quality indicators (i.e., pH, Dissolved Oxygen, Specific Conductance, and Water Temperature). We also apply the method for Spatial Interpolation (Section~\ref{si}) and Temporal Prediction (Section~\ref{tp}) to forecast water quality indicators at any given location and future time. The performance of the six models is evaluated using RMSE (Table~\ref{tab:rmse}, S-T) and $R^2$ (Table~\ref{tab:r2}, S-T). From our observations, XGBoost excels in predicting all four water quality indicators. This success is attributed to its ability to effectively capture the interrelationships between variables and to reveal hidden patterns in the data. XGBoost, with its ensemble learning approach based on gradient-boosted decision trees, comparatively good at capturing the nonlinear spatio-temporal patterns of water quality indicators. This becomes particularly significant for real-world datasets, such as the CWQD used in this paper, which display some outliers and lack continuous data, like the data missing from water quality stations between 1992-1998 (Fig.~\ref{fig.lifeline}). Thus, the robustness of XGBoost assists in addressing the challenges posed by data noise and gaps.

In addition to comparing the models, we analyze how each model performs when estimating different individual water quality indicators. Through the results (Table~\ref{tab:r2}), it can be observed that water temperature has the highest \(R^2\), whereas the \(R^2\) for pH value is the lowest. This reveals that, comparatively, pH has the lowest interpretability, while water temperature has the highest. All these water quality indicators are affected by the environment, but each variable responds to environmental influences differently due to their inherent characteristics. Water has a relatively high specific heat capacity, making it less susceptible to substantial fluctuations due to environmental changes. Moreover, water has a better thermal conductivity than many other substances, meaning that heat can quickly propagate through it, reducing rapid local temperature changes. Since surface water in nature is not composed only of hydrogen and hydroxide ions but also various cations and anions from rocks, soil, atmosphere, and biological activity, when other substances enter the water, they may react with it, forming acids or bases, thereby altering the pH value. Therefore, compared to the stable water temperature, the volatile and easily influenced pH value will have higher uncertainty and lack of representativity during data collection, leading to higher errors and lower interpretability.

\begin{figure*}
    \centering
    \begin{subfigure}[t]{0.49\textwidth}
        \begin{subfigure}[t]{0.03\textwidth}
        \textbf{a)}
        \end{subfigure}
        \begin{subfigure}[t]{0.96\textwidth}
            \includegraphics[width=\textwidth, valign=t]{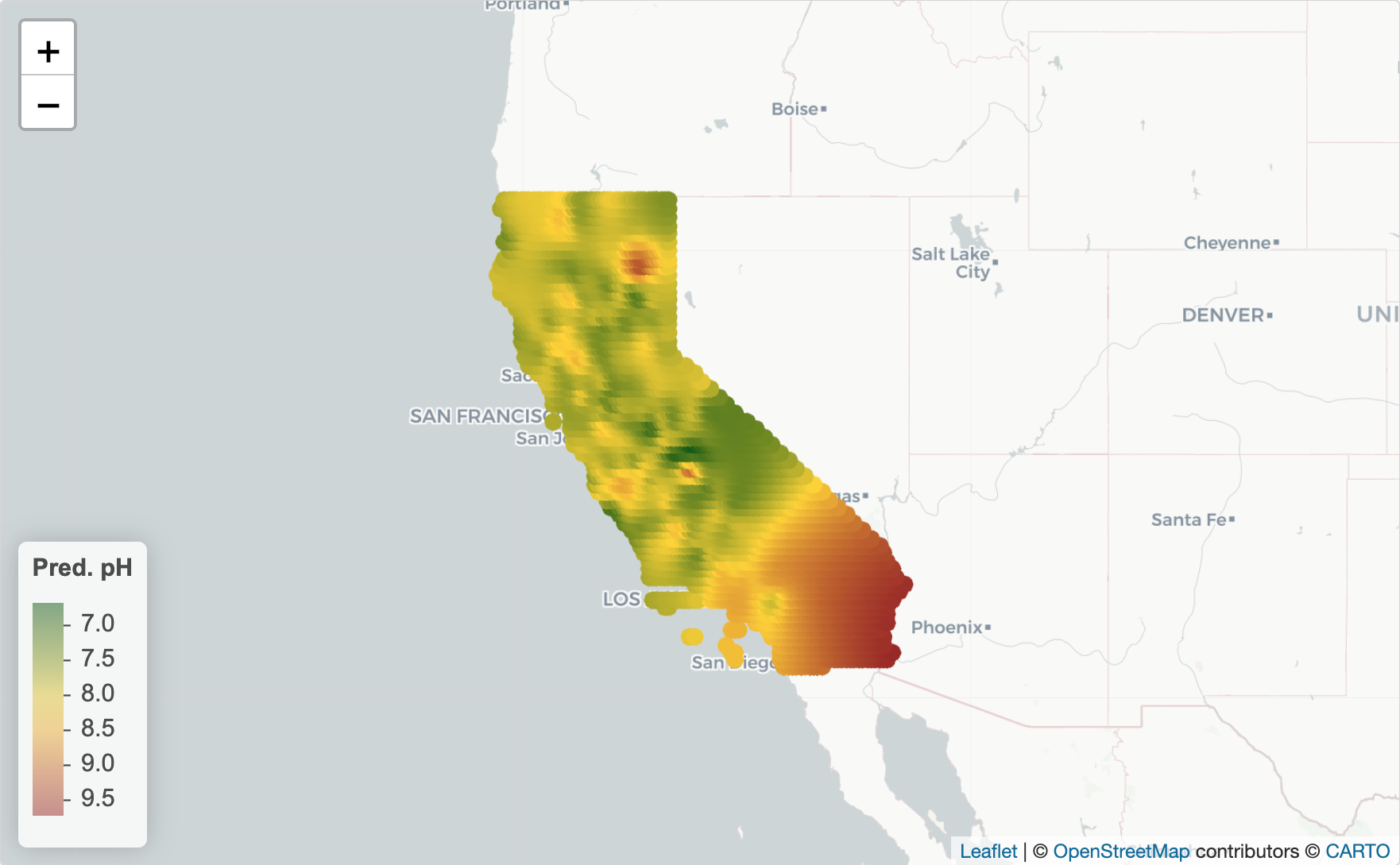}
        \end{subfigure}
    \end{subfigure}
    \hfill
    \begin{subfigure}[t]{0.49\textwidth}
        \begin{subfigure}[t]{0.03\textwidth}
        \textbf{b)}
        \end{subfigure}
        \begin{subfigure}[t]{0.96\textwidth}
            \centering
            \includegraphics[width=\textwidth, valign=t]{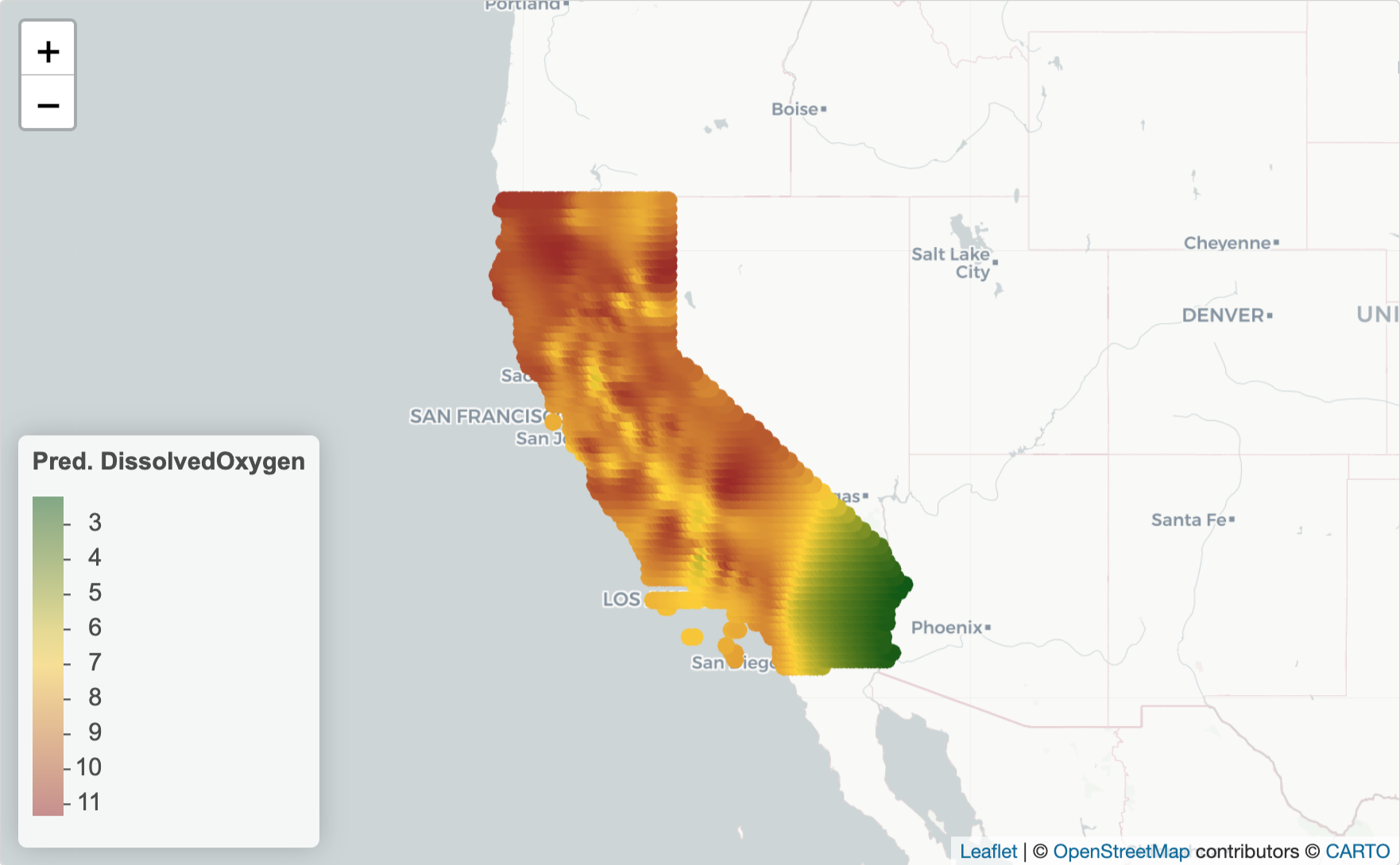}
        \end{subfigure}
    \end{subfigure}    
        \\
    \begin{subfigure}[t]{0.49\textwidth}
        \begin{subfigure}[t]{0.03\textwidth}
        \textbf{c)}
        \end{subfigure}
        \begin{subfigure}[t]{0.96\textwidth}
            \includegraphics[width=\textwidth, valign=t]{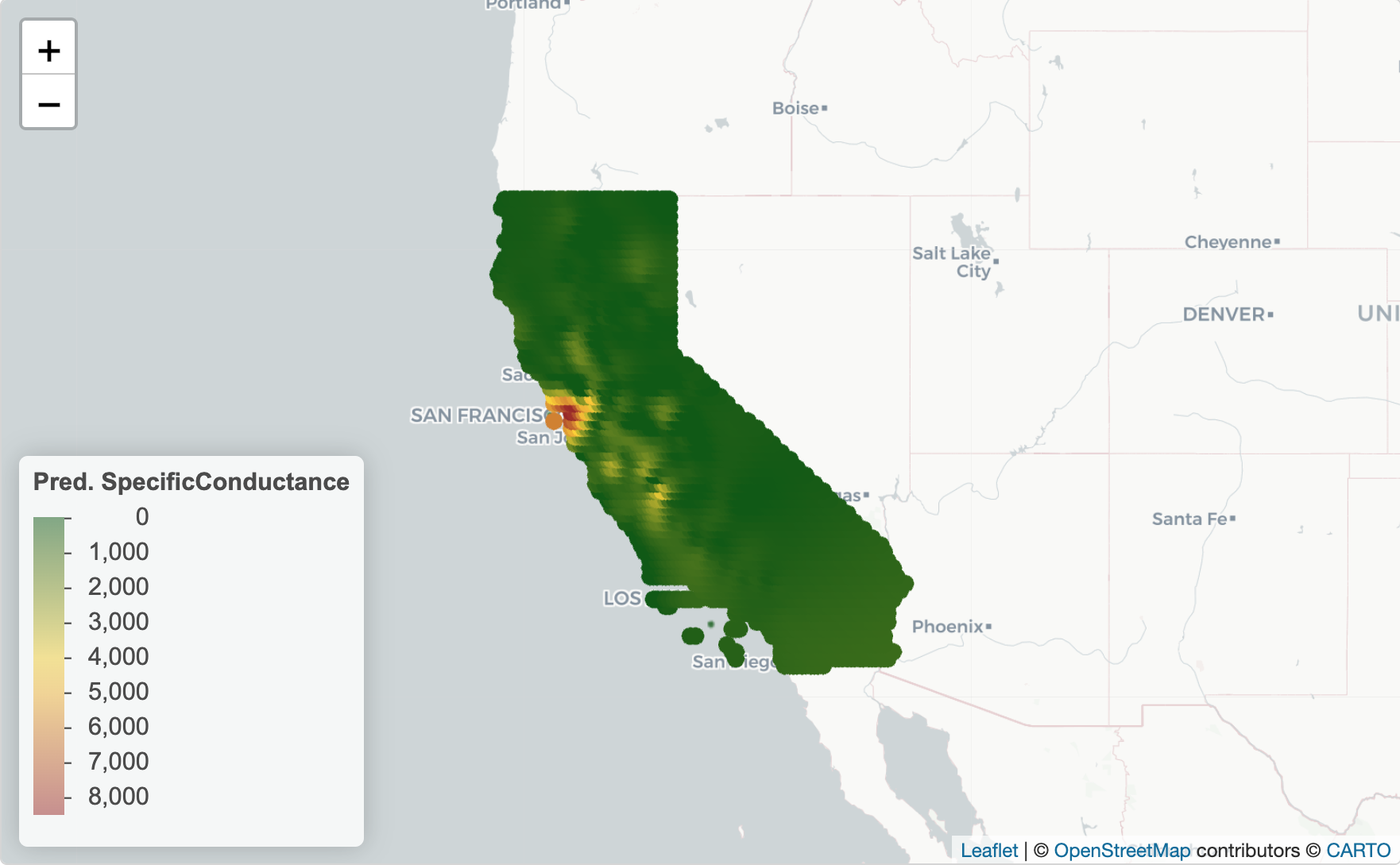}
        \end{subfigure}
    \end{subfigure}
    \hfill
    \begin{subfigure}[t]{0.49\textwidth}
        \begin{subfigure}[t]{0.03\textwidth}
        \textbf{d)}
        \end{subfigure}
        \begin{subfigure}[t]{0.96\textwidth}
            \centering
            \includegraphics[width=\textwidth, valign=t]{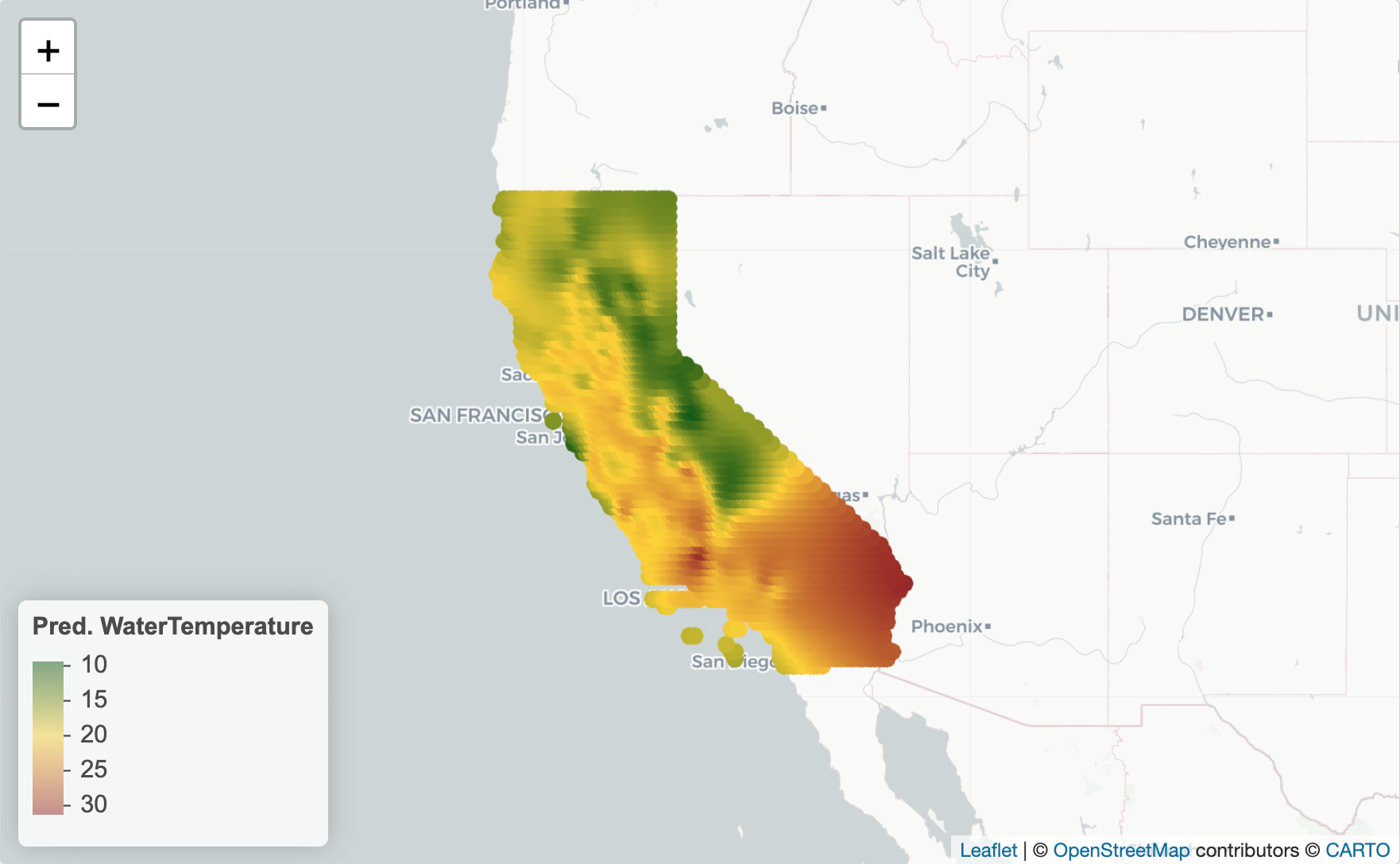}
        \end{subfigure}   
    \end{subfigure}
    \caption{Spatial interpolation for predicting four water quality indicators in July 2023, including pH (a), Dissolved Oxygen (b), Specific Conductance (c), and Water Temperature (d).}
    \label{fig:Interpolation}
\end{figure*}

\subsection{Variable-Dependent Inference} 
\label{v-d}
In addition to Spatio-Temporal Estimations, we also attempt to improve the performance of our model by using prior knowledge such as topography, environmental policies, and other relevant information, and a more detailed analysis of samples based on feature combinations, which will help us identify hidden patterns. Moreover, we aim to use alternative and supplementary variables for forecasting to enhance the predictive accuracy of our target variables and reduce detection costs. It is imperative to acknowledge the intricate interplay and cascading effects among these water quality indicators. For example, as the temperature of the water rises, its capacity to hold dissolved oxygen typically decreases. Subsequently, under low oxygen conditions, certain microorganisms might undergo anaerobic respiration, potentially producing acidic substances such as hydrogen sulfide, leading to a drop in pH. A decrease in pH can instigate the precipitation or dissociation of specific ions, affecting the conductivity of the water. An increase in the electrical conductivity of water could indicate a corresponding increase in the water temperature, as it implies faster ion mobility. Our models are also able to identify and rectify anomalies stemming from operational errors during manual data collection, environmental interference, and the singularity induced by regional constraints. Therefore, we present variable-dependent features for inference, which include three additional water quality indicators, as well as Climate Zone and Geographical Type as categorical variables.

From the RMSE results (Table~\ref{tab:rmse}) and $R^2$ (Table~\ref{tab:r2}), due to errors in manual detection and recording of results, and the interference caused by environmental factors, a significant error margin is evident. For example, anomalies arise, such as water temperatures reaching thousands of degrees, pH values exceeding one hundred, and inconsistent data for the same variable at identical times and locations. The data preprocessing we undertake, coupled with the model's inherent robustness against noise, ensures that the model's results fall within an acceptable range. Consequently, by solely utilizing time, space, and four variables as inputs in Spatio-Temporal Estimation, the models only achieve acceptable performance. However, incorporating other water quality indicators, Climate Zone, and Geographical Type as additional variables significantly enhances the predictive accuracy of the models. Furthermore, these additional variables help identify and rectify anomalies that arise from operational errors during manual data collection, environmental interference, and unique challenges posed by regional constraints.

\begin{figure*}
    \centering
    \begin{subfigure}[t]{0.49\textwidth}
        \begin{subfigure}[t]{0.03\textwidth}
        \textbf{a)}
        \end{subfigure}
        \begin{subfigure}[t]{0.96\textwidth}
            \includegraphics[width=\textwidth, valign=t]{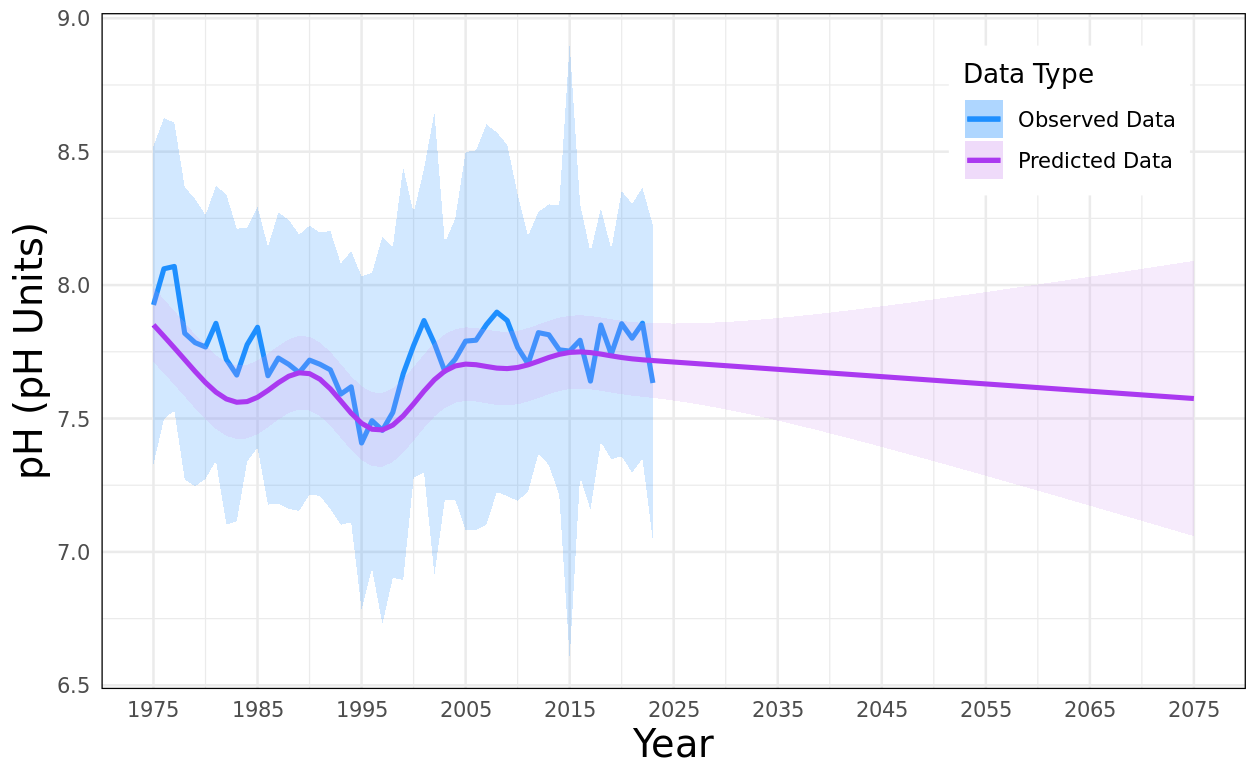}
        \end{subfigure}
    \end{subfigure}
    \hfill
    \begin{subfigure}[t]{0.49\textwidth}
        \begin{subfigure}[t]{0.03\textwidth}
        \textbf{b)}
        \end{subfigure}
        \begin{subfigure}[t]{0.96\textwidth}
            \centering
            \includegraphics[width=\textwidth, valign=t]{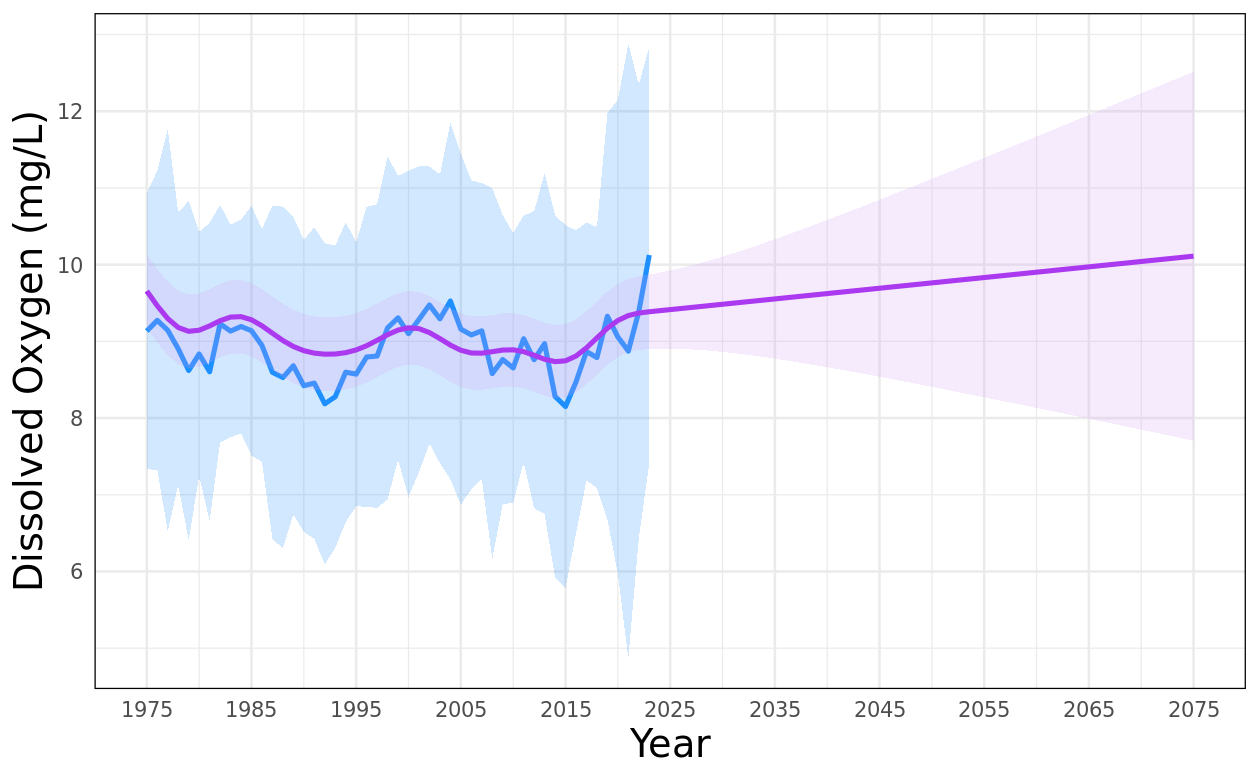}
        \end{subfigure}
    \end{subfigure}    
        \\
    \begin{subfigure}[t]{0.49\textwidth}
        \begin{subfigure}[t]{0.03\textwidth}
        \textbf{c)}
        \end{subfigure}
        \begin{subfigure}[t]{0.96\textwidth}
            \includegraphics[width=\textwidth, valign=t]{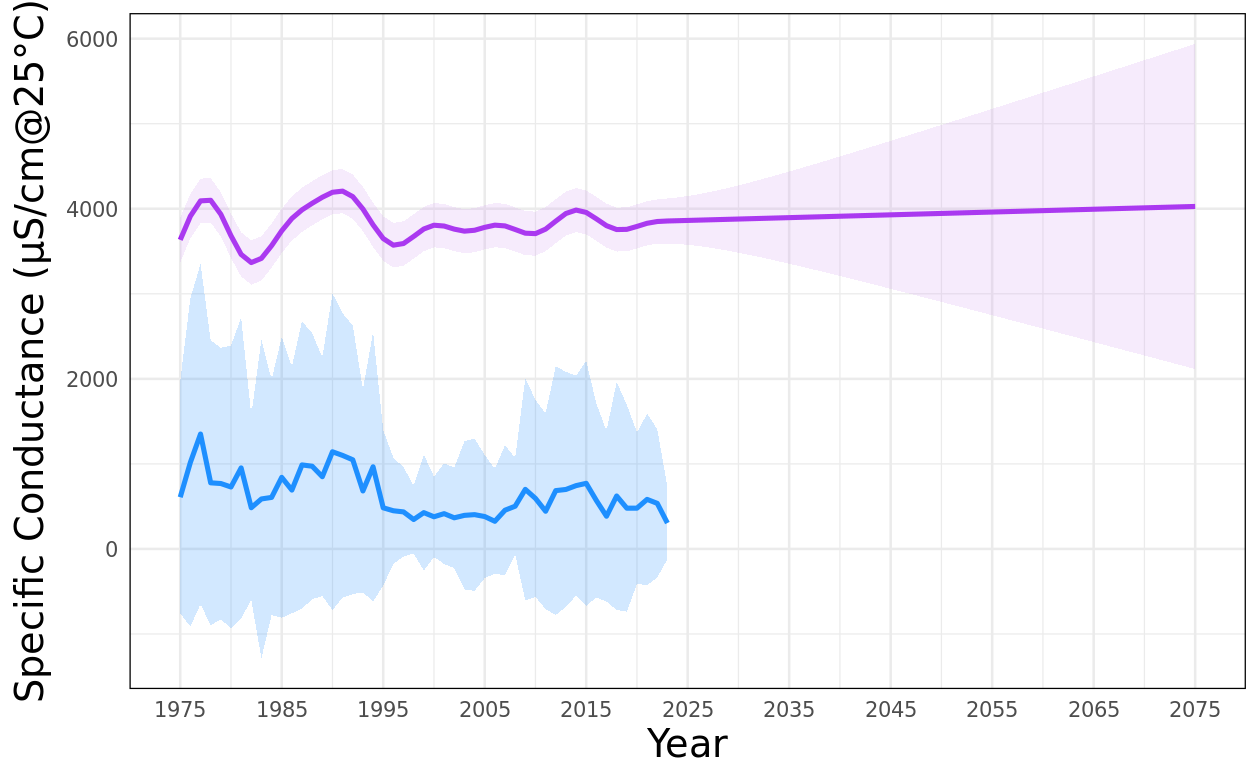}
        \end{subfigure}
    \end{subfigure}
    \hfill
    \begin{subfigure}[t]{0.49\textwidth}
        \begin{subfigure}[t]{0.03\textwidth}
        \textbf{d)}
        \end{subfigure}
        \begin{subfigure}[t]{0.96\textwidth}
            \centering
            \includegraphics[width=\textwidth, valign=t]{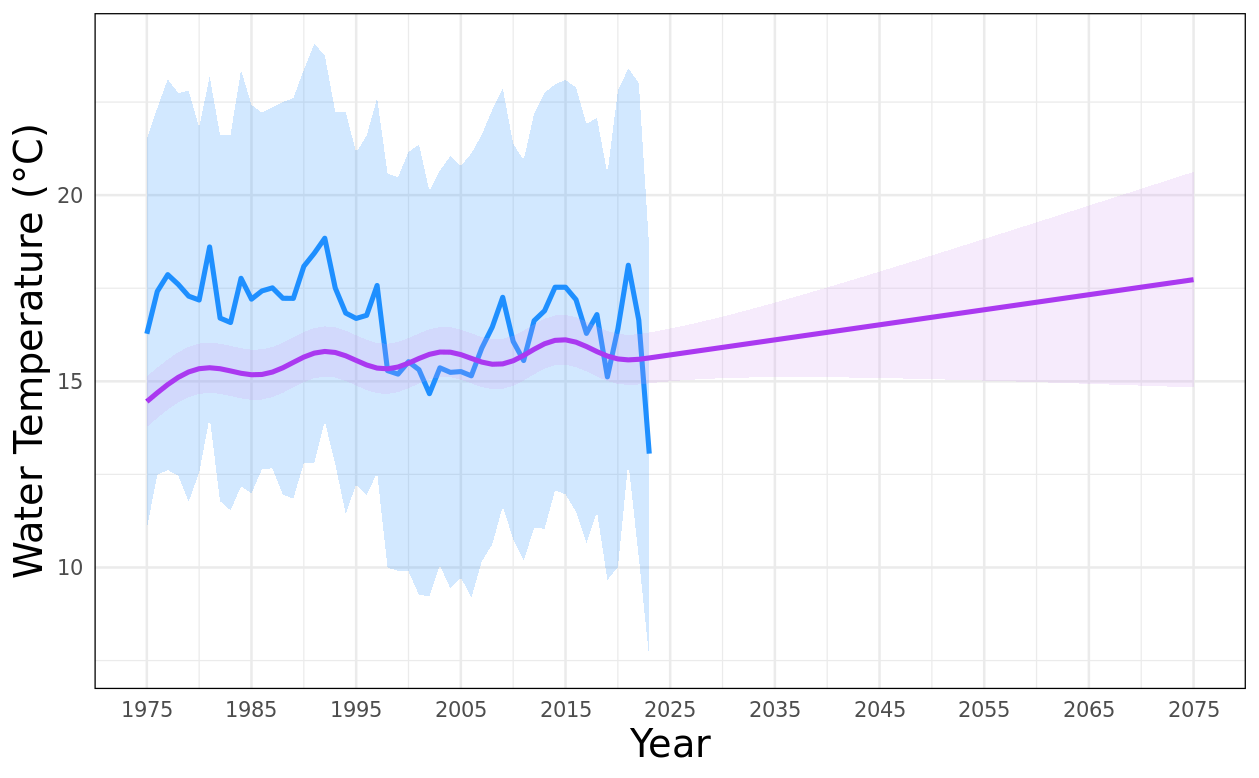}
        \end{subfigure}   
    \end{subfigure}
    \caption{Temporal prediction of the projected changes in four water quality indicators, including pH (a), Dissolved Oxygen (b), Specific Conductance (c), and Water Temperature (d), from 1975 to 2070 at San Francisco (\(37.7749^\circ \text{N}, 122.4194^\circ \text{W}\)). The \textcolor{deepskyblue}{blue line} represents the average values across all water quality stations, while the \textcolor{purpleline}{purple line} denotes the predicted water quality specific to San Francisco. The shaded area signifies the 95\% confidence interval.}
    \label{fig:Temporal}
\end{figure*}

\subsection{Spatial Interpolation}
\label{si}

In this paper, we employ spatio-temporal models to delineate the spatial distribution and trends of four water quality indicators in California. We use the GAM model to estimate the distribution of water quality indicators in California for July 2023 as an example and demonstrate distribution maps of water quality indicators (Fig.~\ref{fig:Interpolation}). These indicators vary spatially but exhibit some intrinsic connections. For example, pH, Dissolved Oxygen, and Water Temperature all demonstrate a gradient variation from north to south, with specific characteristics in the desert region of southern California.

The overall pH values of California's surface water tend to be slightly alkaline. The desert region of southern California displays an approximately uniform pH of around 9, while other areas show sporadic and irregular distributions (Fig.~\ref{fig:Interpolation}a). It is important to note that pH is an indicator of the acidity and alkalinity of water quality, strongly related to the temperature at the time of measurement. In particular, the CWQD, which we use in this paper, is measured and recorded on-site by the experimenter. Therefore, when we estimate surface water pH in July, the field results data may be affected by the increase in pH caused by the overall increase in water temperature in summer. Moreover, the scattered pH values in the northern and central parts of California could be due to a mix of geographical factors and human activities, such as increased aquatic biological activity during summer or irrigation practices in agricultural zones. For the desert region of southern California, the elevated pH might be attributed to evaporation effects and the influence of soil salinity.

Similarly, Dissolved Oxygen follows the same gradient trend as pH (Fig.~\ref{fig:Interpolation}b). This could be attributed to temperature variations, as cooler waters tend to have higher Dissolved Oxygen levels. On the other hand, geographical features, such as dense vegetation in mountainous and forested regions, might contribute to higher Dissolved Oxygen levels in surface waters. Notably, a narrow region in the Central Valley displays lower Dissolved Oxygen levels, which could be due to ecological factors like upwelling or human activities that impact water quality. Generally, most parts of California exhibit relatively high Dissolved Oxygen levels around 10 $mg/L$, indicating a favorable aquatic environment.

For Specific Conductance in California's surface water, there is a consistent pattern, except for anomalously high values observed near the San Francisco Bay Area (Fig.\ref{fig:Interpolation}c). This might be due to the colder California Current from the Pacific Ocean leading to lower ion mobility. Natural filtering and dilution processes in the Central Valley and mountainous areas contribute further to reducing conductivity. However, the elevated Specific Conductance levels near the Bay Area could be linked to industrial and urban outputs, combined with its geographical positioning that experiences erosion from seawater. We examine further the reasons for the extremely high Specific Conductance levels of the San Francisco Bay Area in the Temporal Prediction section (Section~\ref{tp}).

Finally, Water Temperature showcases an evident north-to-south rising trend, with a clear relationship to topography (Fig.~\ref{fig:Interpolation}d). In areas near the desert in southern California, due to the lower latitude and desert climate, Water Temperature can reach a maximum of about \(30 ^\circ \text{C}\). Coastal zones in central California, like the San Francisco Bay Area and Los Angeles, show elevated Water Temperatures possibly due to human activities. In contrast, the forested regions of northern California and the elevated areas of the Sierra Nevada have cooler water temperatures around \(10 ^\circ \text{C}\). This variation can be attributed to their higher latitudes, altitudes, dense forest covers, and melting snow from the mountains. Analyzing water temperatures can also provide insights into climate change, geothermal activities, and vegetation changes.

\subsection{Temporal Prediction}
\label{tp}
To elucidate the reasons behind the extremely high Specific Conductance levels in the San Francisco Bay Area, we employ the GAM model to illustrate the changing trends of four water quality indicators in San Francisco (\(37.7749^\circ \text{N}, 122.4194^\circ \text{W}\)) over a span of 100 years, from 1975 to 2075 (Fig.~\ref{fig:Temporal}). San Francisco's coastal location, port activity, and high population density provide a distinctive contrast to the extensive and diverse terrains of California. The confluence of seawater and freshwater in San Francisco creates a complex milieu for water quality assessment in the Bay Area, influenced by factors such as seawater intrusion, freshwater influx from the Sacramento River and San Joaquin River, and precipitation impacts (\cite{cloern2012drivers, sutton2016microplastic, schlegel2015riverine}).

The pH level in the surface waters of San Francisco follows a trend similar to that observed throughout California (Fig.~\ref{fig:Temporal}a). Statewide, pH levels exhibit significant fluctuations from 1975 to 2023 but tend to remain within a range of 7.5 to 8.5. These fluctuations may be attributable to seasonal variations, climatic shifts, and industrial emissions. As San Francisco transitions from an industrial hub to a nexus of finance, technology, and tourism, model forecasts predict a gradual decline in surface water pH over the forthcoming 50 years, maintaining within a stable range.

The Dissolved Oxygen content in San Francisco's surface water also reflects the overall California trend (Fig.~\ref{fig:Temporal}b). The spatial distribution of Dissolved Oxygen and pH levels are similar, while their temporal trends are inversely related. Dissolved Oxygen levels fluctuate but generally remain between 8 and 11 \(mg/L\), indicating an environment conducive to the maintenance of aquatic life. Projections based on our model suggest a slow increase in Dissolved Oxygen over the next half century, implying an improving aquatic ecosystem in the area.

\begin{figure}
    \centering
    \includegraphics[width=1\linewidth]{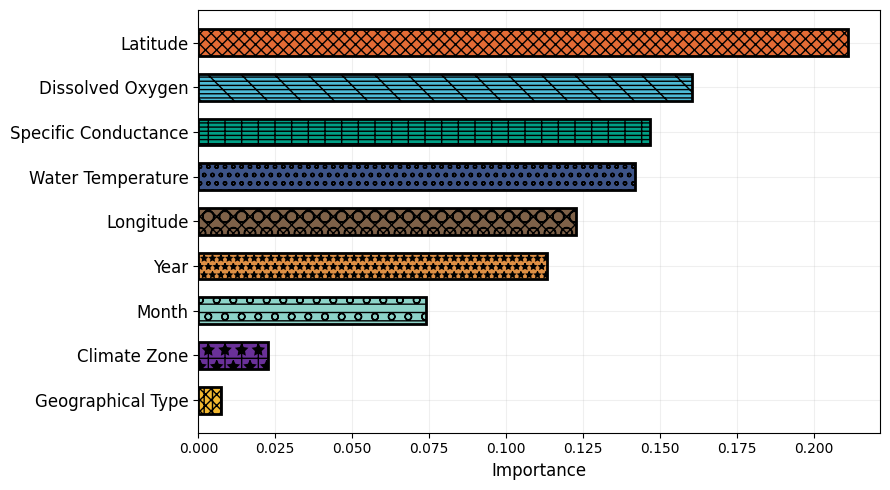}
    \caption{Interpretability of feature variable importances for predicting pH value. The interpretability are exemplified by XGBoost.}
    \label{fig:importance}
\end{figure}

Specific Conductance of San Francisco is significantly higher than the average of California, which reflects a higher ion concentration in the surface water of San Francisco (Fig.~\ref{fig:Temporal}c). This high value might indicate the influence of seawater or pollution. Meanwhile, fluctuations in conductance could result from environmental policies, urban development, and industrial changes. Still, over a span of 100 years, the Specific Conductance does not show a clear upward or downward trend. \cite{work2017record} noted a similar phenomenon from 1990 to 2015, where the mean specific conductance ranged from 10,000 to 50,000 \(\mu S/\text{cm}@25 ^\circ \text{C}\), closely approaching the specific conductance levels found in seawater. Therefore, we speculate that most of the water quality monitoring stations in San Francisco measure surface waters influenced by seawater intrusion. Furthermore, it potentially provides insight into the extent of seawater's reach into the continent.

Water Temperature of San Francisco differs from the overall trend seen across California (Fig.~\ref{fig:Temporal}d). Compared to the erratic fluctuations in the California's Water Temperature, San Francisco displays regular fluctuations with a clear long-term upward trend. This might be because California has varied climates, while San Francisco experiences the temperate (C) climate, where water temperature largely depends on ocean currents. Predicted temperature rises relate to global climate change, particularly rising sea levels and ocean warming. These shifts could affect the ecosystem, the fishing industry, and the economic activities related to tourism in San Francisco.

\subsection{Interpretability of Feature}
Examplified by predicting the pH levels of surface water using the XGBoost model, we present the proportional impact of nine feature variables on the predictions of the model (Fig.~\ref{fig:importance}). It is evident that Latitude holds the highest importance, followed by other three water quality indicators, Longitude, and the two temporal variables, with the two categorical variables (i.e., Climate Zone and Geographical Type) performing the least. There is a significant disparity in the pH values of water across different latitudes in California, making latitude a relatively crucial factor in estimating pH values. This is particularly noticeable in California, given its elongated rectangular shape, where areas of higher Latitude are characterized by extensive forests and mountains, while the lower latitude southern regions are predominantly arid and desert landscapes. Thus, Latitude significantly influences the pH value estimations. Additionally, the intrinsic relationships between water quality indicators also play a vital role, which aligns with our inference in Variable-Dependent Inference (Section~\ref{v-d}). In contrast, the impact of Longitude on pH values is approximately half that of Latitude, which could be attributed to the relatively uniform distribution of climate, water temperature, and topography across California’s longitudinal extent. As for the other two temporal variables, Month and Year, they play a secondary role in pH estimation, indicating a relatively weak correlation with pH values over extended periods like 70 years and across seasonal variations.

\section{Discussion}
The limitations of our study include the suboptimal performance of the model in predicting pH levels. pH is a unitless chemical parameter that measures the logarithmic activity of hydrogen ions and is sensitive to temperature. The inherent variability and instability of pH in natural surface water make it a challenging parameter to predict over the long term. Although the results from the localized areas suggest stable trends, such as the consistent pH values around 8.5 in desert regions influenced by terrain and latitude (Fig.~\ref{fig:Interpolation}a), the overall data across all stations do not exhibit clear overarching temporal trends (Fig.~\ref{fig:Temporal}a). Recent research has revealed that the challenge of pH estimation is due to the active and easily disturbed characteristics of pH in natural surface water. \cite{rosecrans2017predicted} employed a Boosted Regression Trees (BRT) model with more than 60 input variables, the model achieves a RMSE of 0.37 and an \(R^2\) of 0.43. It can be find that, despite considering the uniform terrain, geomorphology, and climate characteristics of the Central Valley, predicting pH values with a large number of input features remains a formidable task. Comparatively, our model encompasses the entire state of California, incorporating various terrains, various climatic conditions, and both Coastal and Inland areas. Moreover, while working with a longer length-of-time dataset (ours 70 years vs. its 21 years) and significantly fewer features (ours 9 vs. its 60+), and our $R^{2}$ value is 10\% higher.

In the process of conducting Variable-Dependent Inference, a series of experiments reveal that categorizing climate into three major climates (B, C, D) yields slightly better performance than subdividing it into nine sub-climates (BWh, BWk, BSh, BSk, Csa, Csb, Dsa, Dsb, Dsc). Taking the prediction of pH values using XGBoost as an example, the former shows an approximate 5\% performance improvement over the latter. The main reason for this phenomenon is the discrepancy in climate indices and time spans between the Köppen climate classification dataset and CWQD. The Köppen dataset is derived from seven other datasets related to temperature, climate, and precipitation, and it relies on different climate indices (\cite{fick2017worldclim, karger2017climatologies}). The literature labels that the confidence near the boundaries of climate zones is relatively low(\cite{beck2018present}). This is particularly evident in California with complex climate zones, exhibiting increased uncertainty. Additionally, there is a discrepancy in the time spans of the datasets. Specifically, the Köppen dataset spans 30 years from 1980 to 2016, while our water quality estimation covers nearly 70 years, from 1956 to 2023. This asymmetry in time spans further exacerbates the uncertainty in the climate classification of water quality stations. Under the influence of global climate warming and related factors, significant changes have been observed in the major and sub-climate classifications in California (\cite{cui2021observed, swain2018increasing, mahlstein2013pace}). Consequently, categorizing a water quality station under a single sub-climate type over an extensive 70-year period appears to be an impractical approach. Although we have introduced the five major climate classification (California only has three major climates) as a method for categorizing the climate at the stations, it has not demonstrated a dominant position in model predictions, possibly due to the impacts of climate change (Fig.~\ref{fig:importance}).

\section{Conclusion}
In this paper, we extensively evaluated four water quality indicators across the entirety of California over the past 70 years, considering both Climate Zone and Geographical Type. We found that the ML model incorporating spatio-temporal variables can reliably estimate four water quality indicators in California, including pH, Dissolved Oxygen, Specific Conductance, and Water Temperature. In addition, the incorporation of other intrinsically related water quality indicators, Climate Zone, and Geographical Type serves to further enhance the performance and robustness of the model. We established interpolations and predictions for four water quality indicators in terms of spatial and temporal dimensions, visualizing their distributions, and projecting their trends for the next 50 years. We observed that the gradient changes of pH, Dissolved Oxygen, and Water Temperature are from north to south. These water quality indicators experienced minor fluctuations in the short term, but the long-term trend is towards a healthier ecological environment. Furthermore, the rise in Water Temperature also signifies the impact of global warming trends on the aquatic ecological environment in the San Francisco Bay Area. We discovered that Specific Conductance is unusually high in the Bay Area, which might indicate the influence of extensive pollution or seawater erosion, potentially lasting up to 100 years. Our future work will dive deeper into other environmental factors that could influence water quality, such as topography, industrial emissions, agricultural activities, and environmental protection policy.

\bibliographystyle{cas-model2-names}

\bibliography{cas-refs}

\begin{thebibliography}{43}
\expandafter\ifx\csname natexlab\endcsname\relax\def\natexlab#1{#1}\fi
\providecommand{\url}[1]{\texttt{#1}}
\providecommand{\href}[2]{#2}
\providecommand{\path}[1]{#1}
\providecommand{\DOIprefix}{doi:}
\providecommand{\ArXivprefix}{arXiv:}
\providecommand{\URLprefix}{URL: }
\providecommand{\Pubmedprefix}{pmid:}
\providecommand{\doi}[1]{\href{http://dx.doi.org/#1}{\path{#1}}}
\providecommand{\Pubmed}[1]{\href{pmid:#1}{\path{#1}}}
\providecommand{\bibinfo}[2]{#2}
\ifx\xfnm\relax \def\xfnm[#1]{\unskip,\space#1}\fi
\bibitem[{Abd-Elaty and Zelenakova(2022)}]{abd2022saltwater}
\bibinfo{author}{Abd-Elaty, I.}, \bibinfo{author}{Zelenakova, M.},
  \bibinfo{year}{2022}.
\newblock \bibinfo{title}{Saltwater intrusion management in shallow and deep
  coastal aquifers for high aridity regions}.
\newblock \bibinfo{journal}{Journal of Hydrology: Regional Studies}
  \bibinfo{volume}{40}, \bibinfo{pages}{101026}.
\bibitem[{Barbieri et~al.(2021)Barbieri, Barberio, Banzato, Billi, Boschetti,
  Franchini, Gori and Petitta}]{barbieri2021climate}
\bibinfo{author}{Barbieri, M.}, \bibinfo{author}{Barberio, M.D.},
  \bibinfo{author}{Banzato, F.}, \bibinfo{author}{Billi, A.},
  \bibinfo{author}{Boschetti, T.}, \bibinfo{author}{Franchini, S.},
  \bibinfo{author}{Gori, F.}, \bibinfo{author}{Petitta, M.},
  \bibinfo{year}{2021}.
\newblock \bibinfo{title}{Climate change and its effect on groundwater
  quality}.
\newblock \bibinfo{journal}{Environmental Geochemistry and Health} ,
  \bibinfo{pages}{1--12}.
\bibitem[{Beck et~al.(2018)Beck, Zimmermann, McVicar, Vergopolan, Berg and
  Wood}]{beck2018present}
\bibinfo{author}{Beck, H.E.}, \bibinfo{author}{Zimmermann, N.E.},
  \bibinfo{author}{McVicar, T.R.}, \bibinfo{author}{Vergopolan, N.},
  \bibinfo{author}{Berg, A.}, \bibinfo{author}{Wood, E.F.},
  \bibinfo{year}{2018}.
\newblock \bibinfo{title}{Present and future k{\"o}ppen-geiger climate
  classification maps at 1-km resolution}.
\newblock \bibinfo{journal}{Scientific data} \bibinfo{volume}{5},
  \bibinfo{pages}{1--12}.
\bibitem[{Brown et~al.(1970)Brown, McClelland, Deininger and
  Tozer}]{brown1970water}
\bibinfo{author}{Brown, R.M.}, \bibinfo{author}{McClelland, N.I.},
  \bibinfo{author}{Deininger, R.A.}, \bibinfo{author}{Tozer, R.G.},
  \bibinfo{year}{1970}.
\newblock \bibinfo{title}{A water quality index-do we dare}.
\newblock \bibinfo{journal}{Water and sewage works} \bibinfo{volume}{117}.
\bibitem[{Buesseler et~al.(2011)Buesseler, Aoyama and
  Fukasawa}]{buesseler2011impacts}
\bibinfo{author}{Buesseler, K.}, \bibinfo{author}{Aoyama, M.},
  \bibinfo{author}{Fukasawa, M.}, \bibinfo{year}{2011}.
\newblock \bibinfo{title}{Impacts of the fukushima nuclear power plants on
  marine radioactivity}.
\newblock \bibinfo{journal}{Environmental science \& technology}
  \bibinfo{volume}{45}, \bibinfo{pages}{9931--9935}.
\bibitem[{{California Department of Water
  Resources}(2023)}]{CaliforniaWaterQualityData}
\bibinfo{author}{{California Department of Water Resources}},
  \bibinfo{year}{2023}.
\newblock \bibinfo{title}{Water quality data}.
\newblock \URLprefix \url{https://data.ca.gov/dataset/water-quality-data}.
  \bibinfo{note}{accessed: 2023-07-31}.
\bibitem[{Chen and Guestrin(2016)}]{chen2016xgboost}
\bibinfo{author}{Chen, T.}, \bibinfo{author}{Guestrin, C.},
  \bibinfo{year}{2016}.
\newblock \bibinfo{title}{Xgboost: A scalable tree boosting system}, in:
  \bibinfo{booktitle}{Proceedings of the 22nd acm sigkdd international
  conference on knowledge discovery and data mining}, pp.
  \bibinfo{pages}{785--794}.
\bibitem[{Chidiac et~al.(2023)Chidiac, El~Najjar, Ouaini, El~Rayess and
  El~Azzi}]{chidiac2023comprehensive}
\bibinfo{author}{Chidiac, S.}, \bibinfo{author}{El~Najjar, P.},
  \bibinfo{author}{Ouaini, N.}, \bibinfo{author}{El~Rayess, Y.},
  \bibinfo{author}{El~Azzi, D.}, \bibinfo{year}{2023}.
\newblock \bibinfo{title}{A comprehensive review of water quality indices
  (wqis): history, models, attempts and perspectives}.
\newblock \bibinfo{journal}{Reviews in Environmental Science and
  Bio/Technology} \bibinfo{volume}{22}, \bibinfo{pages}{349--395}.
\bibitem[{Cloern and Jassby(2012)}]{cloern2012drivers}
\bibinfo{author}{Cloern, J.E.}, \bibinfo{author}{Jassby, A.D.},
  \bibinfo{year}{2012}.
\newblock \bibinfo{title}{Drivers of change in estuarine-coastal ecosystems:
  Discoveries from four decades of study in san francisco bay}.
\newblock \bibinfo{journal}{Reviews of Geophysics} \bibinfo{volume}{50}.
\bibitem[{Cui et~al.(2021)Cui, Liang and Wang}]{cui2021observed}
\bibinfo{author}{Cui, D.}, \bibinfo{author}{Liang, S.}, \bibinfo{author}{Wang,
  D.}, \bibinfo{year}{2021}.
\newblock \bibinfo{title}{Observed and projected changes in global climate
  zones based on k{\"o}ppen climate classification}.
\newblock \bibinfo{journal}{Wiley Interdisciplinary Reviews: Climate Change}
  \bibinfo{volume}{12}, \bibinfo{pages}{e701}.
\bibitem[{Fick and Hijmans(2017)}]{fick2017worldclim}
\bibinfo{author}{Fick, S.E.}, \bibinfo{author}{Hijmans, R.J.},
  \bibinfo{year}{2017}.
\newblock \bibinfo{title}{Worldclim 2: new 1-km spatial resolution climate
  surfaces for global land areas}.
\newblock \bibinfo{journal}{International journal of climatology}
  \bibinfo{volume}{37}, \bibinfo{pages}{4302--4315}.
\bibitem[{Ficklin et~al.(2013)Ficklin, Luo and Zhang}]{ficklin2013watershed}
\bibinfo{author}{Ficklin, D.L.}, \bibinfo{author}{Luo, Y.},
  \bibinfo{author}{Zhang, M.}, \bibinfo{year}{2013}.
\newblock \bibinfo{title}{Watershed modelling of hydrology and water quality in
  the sacramento river watershed, california}.
\newblock \bibinfo{journal}{Hydrological processes} \bibinfo{volume}{27},
  \bibinfo{pages}{236--250}.
\bibitem[{Guinness(2021)}]{guinness2021gaussian}
\bibinfo{author}{Guinness, J.}, \bibinfo{year}{2021}.
\newblock \bibinfo{title}{Gaussian process learning via fisher scoring of
  vecchia’s approximation}.
\newblock \bibinfo{journal}{Statistics and Computing} \bibinfo{volume}{31},
  \bibinfo{pages}{25}.
\bibitem[{Hastie(2017)}]{hastie2017generalized}
\bibinfo{author}{Hastie, T.J.}, \bibinfo{year}{2017}.
\newblock \bibinfo{title}{Generalized additive models}, in:
  \bibinfo{booktitle}{Statistical models in S}. \bibinfo{publisher}{Routledge},
  pp. \bibinfo{pages}{249--307}.
\bibitem[{Hussain et~al.(2019)Hussain, Abd-Elhamid, Javadi and
  Sherif}]{hussain2019management}
\bibinfo{author}{Hussain, M.S.}, \bibinfo{author}{Abd-Elhamid, H.F.},
  \bibinfo{author}{Javadi, A.A.}, \bibinfo{author}{Sherif, M.M.},
  \bibinfo{year}{2019}.
\newblock \bibinfo{title}{Management of seawater intrusion in coastal aquifers:
  a review}.
\newblock \bibinfo{journal}{Water} \bibinfo{volume}{11}, \bibinfo{pages}{2467}.
\bibitem[{Icaga(2007)}]{icaga2007fuzzy}
\bibinfo{author}{Icaga, Y.}, \bibinfo{year}{2007}.
\newblock \bibinfo{title}{Fuzzy evaluation of water quality classification}.
\newblock \bibinfo{journal}{Ecological Indicators} \bibinfo{volume}{7},
  \bibinfo{pages}{710--718}.
\bibitem[{Jha et~al.(2020)Jha, Shekhar and Jenifer}]{jha2020assessing}
\bibinfo{author}{Jha, M.K.}, \bibinfo{author}{Shekhar, A.},
  \bibinfo{author}{Jenifer, M.A.}, \bibinfo{year}{2020}.
\newblock \bibinfo{title}{Assessing groundwater quality for drinking water
  supply using hybrid fuzzy-gis-based water quality index}.
\newblock \bibinfo{journal}{Water Research} \bibinfo{volume}{179},
  \bibinfo{pages}{115867}.
\bibitem[{Karger et~al.(2017)Karger, Conrad, B{\"o}hner, Kawohl, Kreft,
  Soria-Auza, Zimmermann, Linder and Kessler}]{karger2017climatologies}
\bibinfo{author}{Karger, D.N.}, \bibinfo{author}{Conrad, O.},
  \bibinfo{author}{B{\"o}hner, J.}, \bibinfo{author}{Kawohl, T.},
  \bibinfo{author}{Kreft, H.}, \bibinfo{author}{Soria-Auza, R.W.},
  \bibinfo{author}{Zimmermann, N.E.}, \bibinfo{author}{Linder, H.P.},
  \bibinfo{author}{Kessler, M.}, \bibinfo{year}{2017}.
\newblock \bibinfo{title}{Climatologies at high resolution for the earth’s
  land surface areas}.
\newblock \bibinfo{journal}{Scientific data} \bibinfo{volume}{4},
  \bibinfo{pages}{1--20}.
\bibitem[{Kauffman et~al.(2003)}]{kauffman2003climate}
\bibinfo{author}{Kauffman, E.}, et~al., \bibinfo{year}{2003}.
\newblock \bibinfo{title}{Climate and topography}.
\newblock \bibinfo{journal}{Atlas of the Biodiversity of California}
  \bibinfo{volume}{12}, \bibinfo{pages}{15}.
\bibitem[{Kaur et~al.(2010)Kaur, Vats, Rekhi, Bhardwaj, Goel, Tanwar and
  Gaur}]{kaur2010physico}
\bibinfo{author}{Kaur, A.}, \bibinfo{author}{Vats, S.}, \bibinfo{author}{Rekhi,
  S.}, \bibinfo{author}{Bhardwaj, A.}, \bibinfo{author}{Goel, J.},
  \bibinfo{author}{Tanwar, R.S.}, \bibinfo{author}{Gaur, K.K.},
  \bibinfo{year}{2010}.
\newblock \bibinfo{title}{Physico-chemical analysis of the industrial effluents
  and their impact on the soil microflora}.
\newblock \bibinfo{journal}{Procedia Environmental Sciences}
  \bibinfo{volume}{2}, \bibinfo{pages}{595--599}.
\bibitem[{Khadra et~al.(2022)Khadra, Elias and
  Majdalani}]{khadra2022systematic}
\bibinfo{author}{Khadra, W.M.}, \bibinfo{author}{Elias, A.R.},
  \bibinfo{author}{Majdalani, M.A.}, \bibinfo{year}{2022}.
\newblock \bibinfo{title}{A systematic approach to derive natural background
  levels in groundwater: Application to an aquifer in north lebanon perturbed
  by various pollution sources}.
\newblock \bibinfo{journal}{Science of The Total Environment}
  \bibinfo{volume}{847}, \bibinfo{pages}{157586}.
\bibitem[{Kim et~al.(2019)Kim, Jain, Lee, Chen and Park}]{kim2019quantitative}
\bibinfo{author}{Kim, J.S.}, \bibinfo{author}{Jain, S.}, \bibinfo{author}{Lee,
  J.H.}, \bibinfo{author}{Chen, H.}, \bibinfo{author}{Park, S.Y.},
  \bibinfo{year}{2019}.
\newblock \bibinfo{title}{Quantitative vulnerability assessment of water
  quality to extreme drought in a changing climate}.
\newblock \bibinfo{journal}{Ecological Indicators} \bibinfo{volume}{103},
  \bibinfo{pages}{688--697}.
\bibitem[{Li and Zhang(2019)}]{li2019beach}
\bibinfo{author}{Li, J.}, \bibinfo{author}{Zhang, X.}, \bibinfo{year}{2019}.
\newblock \bibinfo{title}{Beach pollution effects on health and productivity in
  california}.
\newblock \bibinfo{journal}{International Journal of Environmental Research and
  Public Health} \bibinfo{volume}{16}, \bibinfo{pages}{1987}.
\bibitem[{Li and Wu(2019)}]{li2019drinking}
\bibinfo{author}{Li, P.}, \bibinfo{author}{Wu, J.}, \bibinfo{year}{2019}.
\newblock \bibinfo{title}{Drinking water quality and public health}.
\newblock \bibinfo{journal}{Exposure and Health} \bibinfo{volume}{11},
  \bibinfo{pages}{73--79}.
\bibitem[{Mahlstein et~al.(2013)Mahlstein, Daniel and
  Solomon}]{mahlstein2013pace}
\bibinfo{author}{Mahlstein, I.}, \bibinfo{author}{Daniel, J.S.},
  \bibinfo{author}{Solomon, S.}, \bibinfo{year}{2013}.
\newblock \bibinfo{title}{Pace of shifts in climate regions increases with
  global temperature}.
\newblock \bibinfo{journal}{Nature Climate Change} \bibinfo{volume}{3},
  \bibinfo{pages}{739--743}.
\bibitem[{Mentaschi et~al.(2018)Mentaschi, Vousdoukas, Pekel, Voukouvalas and
  Feyen}]{mentaschi2018global}
\bibinfo{author}{Mentaschi, L.}, \bibinfo{author}{Vousdoukas, M.I.},
  \bibinfo{author}{Pekel, J.F.}, \bibinfo{author}{Voukouvalas, E.},
  \bibinfo{author}{Feyen, L.}, \bibinfo{year}{2018}.
\newblock \bibinfo{title}{Global long-term observations of coastal erosion and
  accretion}.
\newblock \bibinfo{journal}{Scientific reports} \bibinfo{volume}{8},
  \bibinfo{pages}{12876}.
\bibitem[{Noori et~al.(2019)Noori, Berndtsson, Hosseinzadeh, Adamowski and
  Abyaneh}]{noori2019critical}
\bibinfo{author}{Noori, R.}, \bibinfo{author}{Berndtsson, R.},
  \bibinfo{author}{Hosseinzadeh, M.}, \bibinfo{author}{Adamowski, J.F.},
  \bibinfo{author}{Abyaneh, M.R.}, \bibinfo{year}{2019}.
\newblock \bibinfo{title}{A critical review on the application of the national
  sanitation foundation water quality index}.
\newblock \bibinfo{journal}{Environmental Pollution} \bibinfo{volume}{244},
  \bibinfo{pages}{575--587}.
\bibitem[{Pennino et~al.(2020)Pennino, Leibowitz, Compton, Hill and
  Sabo}]{pennino2020}
\bibinfo{author}{Pennino, M.J.}, \bibinfo{author}{Leibowitz, S.G.},
  \bibinfo{author}{Compton, J.E.}, \bibinfo{author}{Hill, R.A.},
  \bibinfo{author}{Sabo, R.D.}, \bibinfo{year}{2020}.
\newblock \bibinfo{title}{Patterns and predictions of drinking water nitrate
  violations across the conterminous united states}.
\newblock \bibinfo{journal}{Science of the Total Environment}
  \bibinfo{volume}{722}.
\newblock \DOIprefix\doi{10.1016/j.scitotenv.2020.137661}.
\bibitem[{Rosecrans et~al.(2017)Rosecrans, Nolan and
  Gronberg}]{rosecrans2017predicted}
\bibinfo{author}{Rosecrans, C.Z.}, \bibinfo{author}{Nolan, B.T.},
  \bibinfo{author}{Gronberg, J.A.M.}, \bibinfo{year}{2017}.
\newblock \bibinfo{title}{Predicted pH at the domestic and public supply
  drinking water depths, Central Valley, California}.
\newblock \bibinfo{type}{Technical Report}. US Geological Survey.
\bibitem[{Rossi et~al.(2013)Rossi, Van~Sebille, Gupta, Gar{\c{c}}on and
  England}]{rossi2013multi}
\bibinfo{author}{Rossi, V.}, \bibinfo{author}{Van~Sebille, E.},
  \bibinfo{author}{Gupta, A.S.}, \bibinfo{author}{Gar{\c{c}}on, V.},
  \bibinfo{author}{England, M.H.}, \bibinfo{year}{2013}.
\newblock \bibinfo{title}{Multi-decadal projections of surface and interior
  pathways of the fukushima cesium-137 radioactive plume}.
\newblock \bibinfo{journal}{Deep Sea Research Part I: Oceanographic Research
  Papers} \bibinfo{volume}{80}, \bibinfo{pages}{37--46}.
\bibitem[{Sauv{\'e} et~al.(2021)Sauv{\'e}, Lamontagne, Dupras and
  Stahel}]{sauve2021circular}
\bibinfo{author}{Sauv{\'e}, S.}, \bibinfo{author}{Lamontagne, S.},
  \bibinfo{author}{Dupras, J.}, \bibinfo{author}{Stahel, W.},
  \bibinfo{year}{2021}.
\newblock \bibinfo{title}{Circular economy of water: Tackling quantity, quality
  and footprint of water}.
\newblock \bibinfo{journal}{Environmental Development} \bibinfo{volume}{39},
  \bibinfo{pages}{100651}.
\bibitem[{Schlegel and Domagalski(2015)}]{schlegel2015riverine}
\bibinfo{author}{Schlegel, B.}, \bibinfo{author}{Domagalski, J.L.},
  \bibinfo{year}{2015}.
\newblock \bibinfo{title}{Riverine nutrient trends in the sacramento and san
  joaquin basins, california: A comparison to state and regional water quality
  policies}.
\newblock \bibinfo{journal}{San Francisco Estuary and Watershed Science}
  \bibinfo{volume}{13}.
\bibitem[{Sherris et~al.(2021)Sherris, Baiocchi, Fendorf, Luby, Yang and
  Shaw}]{sherris2021}
\bibinfo{author}{Sherris, A.R.}, \bibinfo{author}{Baiocchi, M.},
  \bibinfo{author}{Fendorf, S.}, \bibinfo{author}{Luby, S.P.},
  \bibinfo{author}{Yang, W.}, \bibinfo{author}{Shaw, G.M.},
  \bibinfo{year}{2021}.
\newblock \bibinfo{title}{Nitrate in drinking water during pregnancy and
  spontaneous preterm birth: A retrospective within-mother analysis in
  california}.
\newblock \bibinfo{journal}{Environmental Health Perspectives}
  \bibinfo{volume}{129}.
\newblock \DOIprefix\doi{10.1289/EHP8205}.
\bibitem[{Sutton et~al.(2016)Sutton, Mason, Stanek, Willis-Norton, Wren and
  Box}]{sutton2016microplastic}
\bibinfo{author}{Sutton, R.}, \bibinfo{author}{Mason, S.A.},
  \bibinfo{author}{Stanek, S.K.}, \bibinfo{author}{Willis-Norton, E.},
  \bibinfo{author}{Wren, I.F.}, \bibinfo{author}{Box, C.},
  \bibinfo{year}{2016}.
\newblock \bibinfo{title}{Microplastic contamination in the san francisco bay,
  california, usa}.
\newblock \bibinfo{journal}{Marine pollution bulletin} \bibinfo{volume}{109},
  \bibinfo{pages}{230--235}.
\bibitem[{Swain et~al.(2018)Swain, Langenbrunner, Neelin and
  Hall}]{swain2018increasing}
\bibinfo{author}{Swain, D.L.}, \bibinfo{author}{Langenbrunner, B.},
  \bibinfo{author}{Neelin, J.D.}, \bibinfo{author}{Hall, A.},
  \bibinfo{year}{2018}.
\newblock \bibinfo{title}{Increasing precipitation volatility in
  twenty-first-century california}.
\newblock \bibinfo{journal}{Nature Climate Change} \bibinfo{volume}{8},
  \bibinfo{pages}{427--433}.
\bibitem[{Tahmasebi et~al.(2020)Tahmasebi, Kamrava, Bai and
  Sahimi}]{tahmasebi2020machine}
\bibinfo{author}{Tahmasebi, P.}, \bibinfo{author}{Kamrava, S.},
  \bibinfo{author}{Bai, T.}, \bibinfo{author}{Sahimi, M.},
  \bibinfo{year}{2020}.
\newblock \bibinfo{title}{Machine learning in geo-and environmental sciences:
  From small to large scale}.
\newblock \bibinfo{journal}{Advances in Water Resources} \bibinfo{volume}{142},
  \bibinfo{pages}{103619}.
\bibitem[{Tariqi and Naughton(2021)}]{tariqi2021water}
\bibinfo{author}{Tariqi, A.Q.}, \bibinfo{author}{Naughton, C.C.},
  \bibinfo{year}{2021}.
\newblock \bibinfo{title}{Water, health, and environmental justice in
  california: Geospatial analysis of nitrate contamination and thyroid cancer}.
\newblock \bibinfo{journal}{Environmental Engineering Science}
  \bibinfo{volume}{38}, \bibinfo{pages}{377--388}.
\bibitem[{Uddin et~al.(2021)Uddin, Nash and Olbert}]{Uddin2021}
\bibinfo{author}{Uddin, G.}, \bibinfo{author}{Nash, S.},
  \bibinfo{author}{Olbert, A.I.}, \bibinfo{year}{2021}.
\newblock \bibinfo{title}{{A review of water quality index models and their use
  for assessing surface water quality}}.
\newblock \bibinfo{journal}{Ecological Indicators} \bibinfo{volume}{122},
  \bibinfo{pages}{107218}.
\bibitem[{Work et~al.(2017)Work, Downing-Kunz and Livsey}]{work2017record}
\bibinfo{author}{Work, P.A.}, \bibinfo{author}{Downing-Kunz, M.A.},
  \bibinfo{author}{Livsey, D.N.}, \bibinfo{year}{2017}.
\newblock \bibinfo{title}{Record-high specific conductance and water
  temperature in San Francisco Bay during water year 2015}.
\newblock \bibinfo{type}{Technical Report}. US Geological Survey.
\bibitem[{{World Health Organization}(2022)}]{world2022guidelines}
\bibinfo{author}{{World Health Organization}}, \bibinfo{year}{2022}.
\newblock \bibinfo{title}{Guidelines for drinking-water quality: incorporating
  the first and second addenda}.
\newblock \bibinfo{publisher}{World Health Organization}.
\bibitem[{Yuan et~al.(2023a)Yuan, Chen, Ewing, Blasch and Li}]{yuan2023three}
\bibinfo{author}{Yuan, L.}, \bibinfo{author}{Chen, H.}, \bibinfo{author}{Ewing,
  R.}, \bibinfo{author}{Blasch, E.}, \bibinfo{author}{Li, J.},
  \bibinfo{year}{2023}a.
\newblock \bibinfo{title}{Three dimensional indoor positioning based on passive
  radio frequency signal strength distribution}.
\newblock \bibinfo{journal}{IEEE Internet of Things Journal}
  \bibinfo{volume}{10}, \bibinfo{pages}{13933 -- 13944}.
\bibitem[{Yuan et~al.(2023b)Yuan, Chen, Ewing and Li}]{yuan2023passive}
\bibinfo{author}{Yuan, L.}, \bibinfo{author}{Chen, H.}, \bibinfo{author}{Ewing,
  R.}, \bibinfo{author}{Li, J.}, \bibinfo{year}{2023}b.
\newblock \bibinfo{title}{Passive radio frequency-based 3d indoor positioning
  system via ensemble learning}.
\newblock \bibinfo{journal}{arXiv preprint arXiv:2304.06513} .
\bibitem[{Zhu et~al.(2022)Zhu, Wang, Yang, Zhang, Zhang, Ren, Wu and
  Ye}]{zhu2022review}
\bibinfo{author}{Zhu, M.}, \bibinfo{author}{Wang, J.}, \bibinfo{author}{Yang,
  X.}, \bibinfo{author}{Zhang, Y.}, \bibinfo{author}{Zhang, L.},
  \bibinfo{author}{Ren, H.}, \bibinfo{author}{Wu, B.}, \bibinfo{author}{Ye,
  L.}, \bibinfo{year}{2022}.
\newblock \bibinfo{title}{A review of the application of machine learning in
  water quality evaluation}.
\newblock \bibinfo{journal}{Eco-Environment \& Health} .

\end{thebibliography}




\end{document}